\documentclass[11pt]{article}
\usepackage[utf8]{inputenc}
\usepackage{jheppub}

\usepackage{epsfig,latexsym,cancel}
\usepackage{amssymb,amsmath}
\usepackage{epstopdf}
\usepackage{hyperref, graphicx, slashed, xcolor, subfigure}
\usepackage{multirow}

\usepackage{tikz}
\usepackage{tikz-feynman}
\tikzfeynmanset{compat=1.0.0}
\pgfdeclarelayer{bg}    
\pgfsetlayers{bg,main}  

\definecolor{forestgreen}{HTML}{228B22}

 \usepackage{comment}

\newcommand{\brr}[1]{\left(#1\right)}
\newcommand{\srr}[1]{\left[#1\right]}

\newcommand{\beq}{\begin{equation}}
\newcommand{\eeq}{\end{equation}}
\newcommand{\bea}{\begin{eqnarray}}
\newcommand{\eea}{\end{eqnarray}}

\newcommand{\abs}[1]{\left\lvert #1 \right\rvert}

\newcommand{\hyphen}{\,\mathchar`-\mathchar`-\,}

\title{Flavor-changing light bosons with accidental longevity}

\author[a,b]{Yohei Ema,}
\author[b]{Zhen Liu,}
\author[b]{Kun-Feng Lyu,}
\author[a,b]{Maxim Pospelov}
\affiliation[a]{William I. Fine Theoretical Physics Institute, School of Physics and Astronomy,
University of Minnesota, Minneapolis, M 55455, USA}
\affiliation[b]{School of Physics and Astronomy, University of Minnesota, Minneapolis, MN 55455, USA}

\emailAdd{ ema00001@umn.edu}
\emailAdd{ lyu00145@umn.edu}
\emailAdd{ zliuphys@umn.edu}
\emailAdd{ pospelov@umn.edu}

\preprint{UMN-TH-4205/22 \\   
\rightline{FTPI-MINN-22-30} }

\abstract{
We consider a model with a complex scalar field that couples to $(e,\mu)$ or $(\mu,\tau)$ within the ``longevity" window: $[|m_{l_1} - m_{l_2}|, m_{l_1} + m_{l_2}]$ in which $l_1$ and $l_2$ are the two different charged leptons. Within such a mass window, even a relatively large coupling ({\em e.g.} of the size commensurate with the current accuracy/discrepancy in the muon $g-2$ experiment) leads to long lifetimes and macroscopic propagation distance between production and decay points. 
We propose to exploit several existing neutrino experiments and one future experiment to probe the parameter space of this model. For the $\mu-e$ sector, we exploit the muonium decay branching ratio and the production and decay sequence at the LSND experiment, excluding the parametric region suggested by $g_\mu -2$ anomaly. For the $\tau-\mu$ sector, we analyze three main production mechanisms of scalars at beam dump experiments: the Drell-Yan process, the heavy meson decay, and the muon scattering. We explore the constraints from the past CHARM and NuTeV
experiments, and evaluate sensitivity for the proposed beam dump experiment, SHiP. 
The latter can thoroughly probe the parameter space relevant for the $g_\mu - 2$ anomaly.
}

\begin{document}

\maketitle

\section{Introduction}

Lepton flavor is an accidental symmetry which is exactly conserved in the SM with massless neutrinos. The observation of neutrino oscillation and the non-vanishing neutrino mass have firmly established the existence of neutral lepton flavor transitions. This provides strong motivation to search for the charged lepton flavor violation (CLFV). In the Standard Model (SM), the CLFV is suppressed by $G_F^2 m_\nu^4 \leq 10^{-50}$ which is tiny and beyond the sensitivity of any experiments. Thus, any observation of CLFV must imply new physics beyond Standard Model.  There has been a great interest in searching CLFV in various experiments (see recent review like~\cite{Davidson:2022jai}).  In the next decades, the experimental search can be greatly improved by the new generation of experiments such as MEG II, Mu3e, COMET and Mu2e~\cite{Meucci:2022qbh,Chiappini:2021ytk,Mu2e:2014fns,Hesketh:2022wgw,COMET:2018auw}. 
It is also expected that Belle-II~\cite{Belle-II:2018jsg} experiment will collect around $5 \times 10^{10}$ events with $\tau^\pm$ pairs, which is one order enhancement compared to Belle~\cite{Belle:2012iwr} and BaBar~\cite{BaBar:2001yhh}. Additional support for the search of the charged lepton flavor violation comes from the ongoing discrepancy between the measured and predicted anomalous magnetic moment of the muon, which may indicate presence of New Physics in the muon sector. 

Theoretically, there are various ways to realize the CLFV in the new physics sector at different energy scales. Flavor off-diagonal contribution can originate from heavy degrees of freedom. For example, the flavor non-diagonal SUSY breaking terms, and R-violation terms~\cite{Paradisi:2005fk,Girrbach:2009uy,Hisano:2001qz,Babu:2002et} can generate Lepton Flavor Violation (LFV) terms. For the purpose of low-energy physics, one can integrate these heavy degrees of freedom out and parametrize the LFV in terms of the higher dimensional operators. The complete sets of dimension-six operators can be found in {\em e.g.}~\cite{Crivellin:2013hpa}. This way, precision constraints on branching ratios of multiple rare decay processes can be translated to limits on the corresponding Wilson coefficients in the effective theory.  
An interesting alternative to this approach exists in theories with light beyond SM states that can mediate such flavor transitions. 
This results in a somewhat richer phenomenology, with the possibility of studying the production and decay of such light states using various experimental techniques. 

There are many well-known theoretical models which can provide such kinds of couplings. One well-known example is the axion or axion-like particles (ALP). It was originally postulated to solve the QCD strong CP problem and can be defined as the pseudo-Nambu-goldstone boson from the spontaneous breaking of the Peccei-Quinn symmetry at some high scale $f_a$. Subsequently, the QCD axion has been generalized to an ALP, and its mass could cover a wide range of scales from ultralight sub-eV scales to above the electroweak scale. An ALP can couple to leptons and quarks off-diagonally, thus leading to flavored $l_1-l_2-a$ couplings~\cite{Wilczek:1982rv,Ema:2016ops,Calibbi:2016hwq,Bonnefoy:2020llz,Bauer:2021mvw,MartinCamalich:2020dfe,Calibbi:2020jvd}. 
An exchange of real ALP will then generate 
$l_1l_1\to l_2 l_2$ LFV transitions. 
Similarly, the pseudo-scalar Majoron and familon~\cite{Farzan:2002wx,Kachelriess:2000qc,Lessa:2007up,Wilczek:1982rv,Reiss:1982sq,Chang:1987hz}, which are the Goldstone boson relevant to the breaking of the global lepton number and global family symmetry, can also make contributions to LFV.

In this paper, we will introduce a flavor changing scalar (FCS) 
with the flavor off-diagonal coupling as a possible solution to the on-going discrepancy between SM prediction and measurement of $(g-2)_\mu$~\cite{Muong-2:2002wip,Muong-2:2004fok,Muong-2:2006rrc,Davier:2017zfy,Colangelo:2018mtw,Hoferichter:2019mqg,Davier:2019can,Keshavarzi:2019abf,Aoyama:2020ynm,Muong-2:2021ojo}. 
It is well known that one loop correction to $(g-2)_\mu$ resulting from a ALP loop is typically of a ``wrong" sign of curing the anomaly if ALP has a flavor-diagonal coupling. If, on the other hand, the coupling connects two different flavors, the sign of the correction can be positive, and one could consider such a model as a candidate solution to the muon $(g-2)_\mu$ discrepancy. (One has to remark at this point that the status of the discrepancy was questioned recently by the lattice QCD results, which came much closer to the observed value \cite{Borsanyi:2020mff}.)

Let us consider the $l_1-l_2-\phi$ coupling between an FCS and two leptons ($m_{l_1} < m_{l_2}$) in the range relevant for the $(g-2)_\mu$ correction. 
Then, with respect to a possible mass ordering of $\phi$, one can point to the three main regimes:
\begin{eqnarray}
    m_\phi < m_{l_2}-m_{l_1}, ~~~{\rm prompt~decay~ of}~l_2,\\
    m_{l_2}-m_{l_1}< m_\phi < m_{l_2}+m_{l_1}
    ~~~{\rm longevity~window},\\\
    m_\phi > m_{l_2}+m_{l_1} ~~~{\rm prompt~decay~ of}~\phi.
\end{eqnarray}
The first regime leads to fast decays of $l_2\to l_1 + \phi$, and is generally excluded for the range of coupling of interest by the lifetime of $l_2$, while the third regime leads to the prompt decay of $\phi$. The longevity window is when neither of the two-body decays, $l_2\to l_1 + \phi$ and $\phi \to l_2+ l_1$ are allowed. In this window, $\phi$ is long-lived despite sizable coupling provided that $l_1-l_1-\phi$ coupling is forbidden.

This paper mainly considers a simple FCS model with flavor off-diagonal transitions in the lepton sector. 
A light complex scalar $\phi$ is introduced, which directly couples to $(\mu e)$ or $(\tau \mu)$. Instead of breaking the lepton flavor symmetry at the Lagrangian level, we assign the complex scalar to carry both the charged lepton charge with opposite signs. For the sake of generality, we do not assume the new boson is scalar or pseudoscalar, but the scalar and axial coupling are both turned on.

We show that lepton $g-2$ experiments provide sensitive probe of $l_1-l_2-\phi$  vertex. 
In addition, $\phi$ particle longevity opens an interesting possibility for the beam dump studies of $\phi$. While the off-diagonal $e\mu $ coupling can be efficiently studied at intense sources of muons, the $\mu\tau$ case can be probed directly at the proton beam dump facilities and neutrino experiments. 
We analyze three different mechanisms of $\phi$ production: electroweak production in the target, production through fragmentation of charged mesons, and secondary muon-initiated production. 
We then exploit the results of the existing experiments (LSND~\cite{LSND:1996jxj,LSND:2001akn}, NuTeV~\cite{NuTeV:2001ndo,NuTeV:2001ndo,NuTeV:2005wsg} and CHARM~\cite{CHARM:1985nku}), to set competitive constraints on the parameter space of the model. It turns out that using existing experimental results, one can derive direct
constraints on $e-\mu-\phi$ coupling that are better than indirect ones obtained via muon $g-2$. At the same time, indirect constraints on $\mu-\tau-\phi$ coupling are stronger than the direct ones. 
We also derive projections for the SHiP experiment~\cite{Alekhin:2015byh,SHiP:2020hyy}, and show that it can be used to probe the muon $g-2$ relevant window of couplings.

The rest of this paper is organized as follows. In Sec~\ref{sec:preliminary}, we introduce the more detailed model description and establish the connection to the magnetic moment observables. Then we analyze the signatures of the model for a variety of past, current and proposed experiments. We cover the $\mu-e$ case in Sec.~\ref{sec:mu-e}, and $\tau-\mu$ case in Sec.~\ref{sec:tau-mu}. We reach our conclusions in Sec.~\ref{sec:summary}.

\section{Preliminaries}
\label{sec:preliminary}
In this section, we summarize the basic ingredients of our setup.
We describe the model with a complex scalar field $\phi$ in Sec.~\ref{subsec:model}.
We then compute various electromagnetic moments that $\phi$ induces,
with an emphasis on the muon anomalous magnetic moment in Sec.~\ref{subsec:muMDM}.

\subsection{Model description}
\label{subsec:model}
The ongoing discrepancy between the SM prediction and the measured value of the muon $g-2$
motivates an extension of the SM in the muon sector.
In this paper, we consider a complex scalar field $\phi$ that couples to leptons
\begin{align}
	\mathcal{L} &= \abs{\partial \phi}^2 - m_\phi^2 \abs{\phi}^2
 	+ \phi\,\bar{\mu} \left(g_V + g_A\gamma_5 \right)l
	+ \phi^* \bar{l}\left(g_V^* - g_A^*\gamma_5 \right)\mu,
	\label{eq:full_lagrangian}
\end{align}
where $l$ is either $e$ or $\tau$. 
We assume that $\phi$ has $L_\mu$ charge $+1$ and $L_l$ charge $-1$
so that only the couplings with different generations of leptons are allowed.\footnote{
	This charge assignment prevents $\phi$ from meditating, e.g., mixing between muonium and anti-muonium,
	and same-sign four lepton final state production at colliders, in contrast to a real scalar case
	(see e.g.~\cite{Endo:2020mev,Iguro:2020rby}).
}
This Lagrangian may originate from
\begin{align}
	\mathcal{L} &= \abs{\partial \phi}^2 - m_\phi^2 \abs{\phi}^2
	+ \frac{\partial_\mu \phi}{\Lambda} \bar{\mu} \gamma^\mu\left(\tilde{g}_V - \tilde{g}_A \gamma_5\right)l
	+ \frac{\partial_\mu \phi^*}{\Lambda}\bar{l}\gamma^\mu
	\left(\tilde{g}_V^* - \tilde{g}_A^* \gamma_5\right) \mu,
\end{align}
that has only derivative interactions between $\phi$ and SM leptons.
Indeed, after using the equation of motion of the leptons 
we reproduce Eq.~\eqref{eq:full_lagrangian}.
As stated in the introduction, throughout this paper, we focus on the mass range
\begin{align}
	\abs{m_\mu - m_l} < m_\phi < m_\mu + m_l.
\end{align}
Although relatively a limited mass range, 
this window provides interesting phenomenological signatures.
The lower bound prohibits two-body decay $\mu \to \phi + e$ or $\tau \to \phi + \mu$,
avoiding stringent constraints on the couplings,
while the upper bound prohibits two-body decay $\phi \to \mu + l$,
making $\phi$ accidentally long-lived.
This model can explain the discrepancy of the muon anomalous magnetic dipole moment
as we now see.

\subsection{Muon anomalous magnetic moment}
\label{subsec:muMDM}

\begin{figure}[t]
	\centering
	\includegraphics[width=0.475\linewidth]{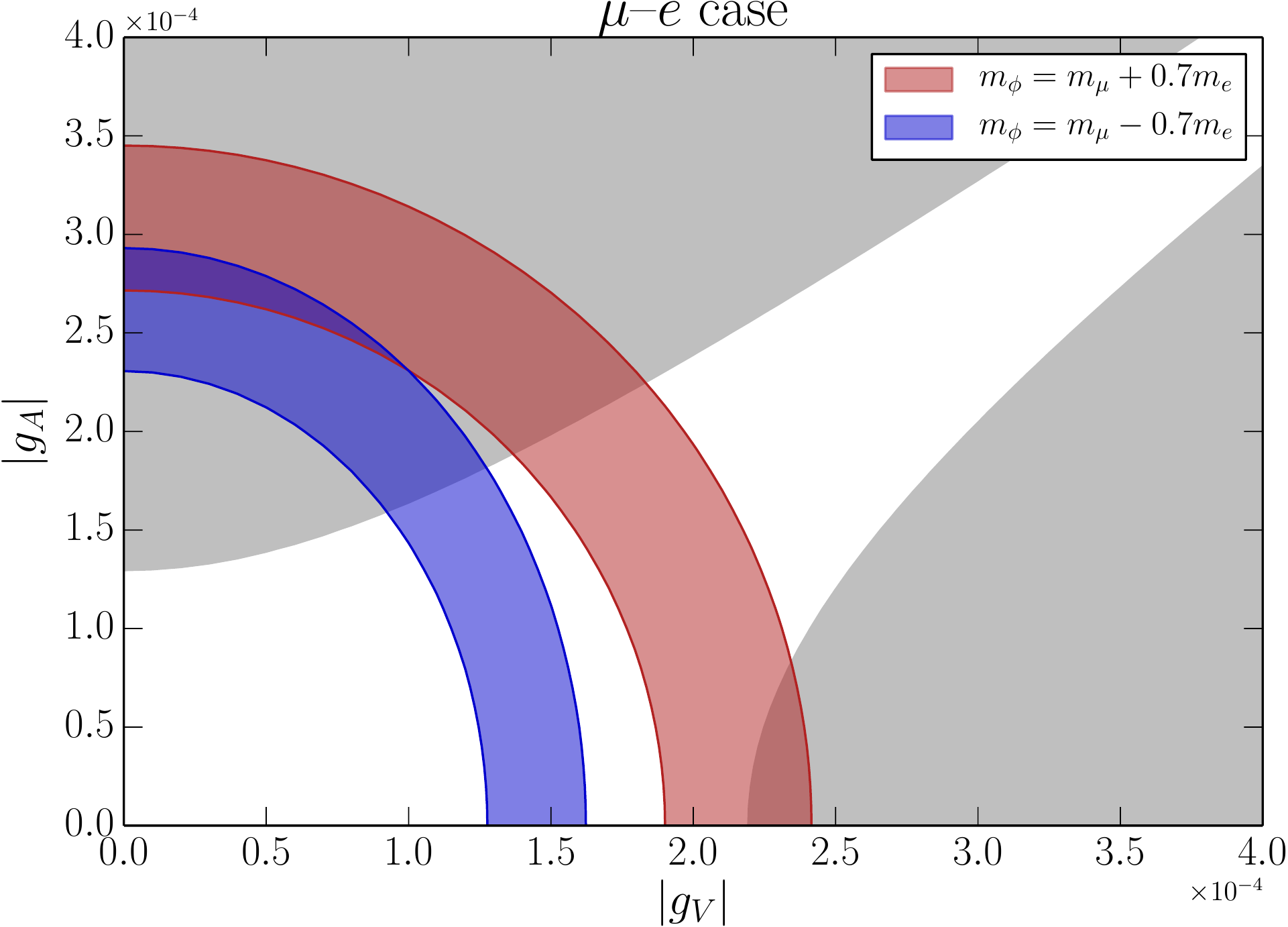}
	\hspace{0mm}
	\includegraphics[width=0.475\linewidth]{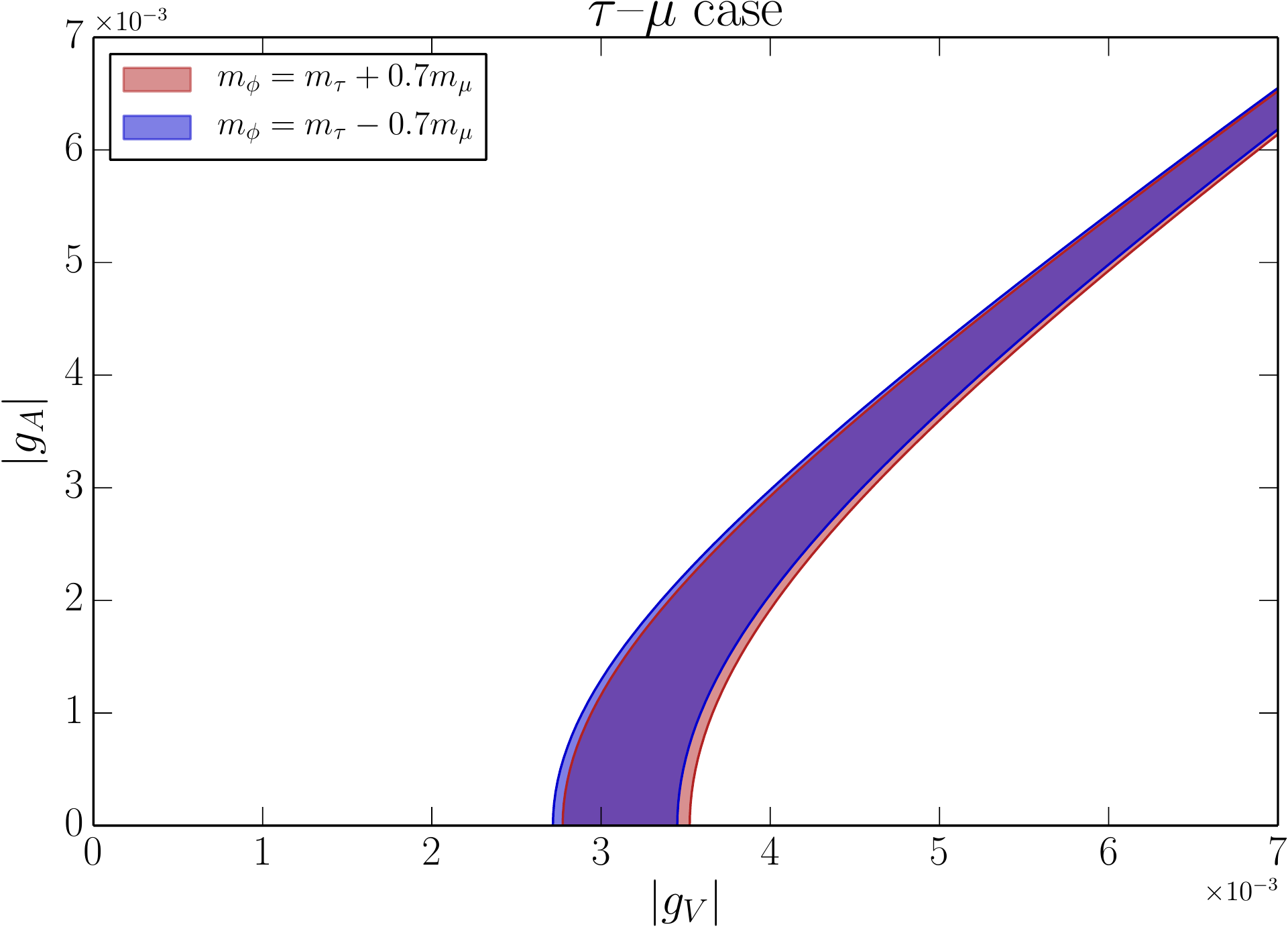}
 	\caption{\small  
	The parameter regions that explain the muon $g-2$ anomaly.
	Note that the magnetic dipole moments depend only on the absolute values of $g_V$ and $g_A$.
	\emph{Left panel:} the $\mu\hyphen e$ case. 
	The red and blue regions explain the muon $g-2$ anomaly
	within $1\sigma$ with $m_\phi = m_\mu + 0.7 m_e$ and $m_\phi = m_\mu - 0.7 m_e$,
	respectively, as benchmarks.
	The gray region is excluded by the electron $g-2$ constraint~\eqref{eq:eMDM}.
	We have taken $m_\phi = m_\mu$ for the electron $g-2$ constraint in this figure,
	but the electron $g-2$ constraint
	has little dependence on $m_\phi$ within the mass range of our interest.
	\emph{Right panel:} the $\tau \hyphen\mu$ case. 
	The red and blue regions explain the muon $g-2$ anomaly
	within $1\sigma$ with $m_\phi = m_\tau + 0.7 m_\mu$ and $m_\phi = m_\tau - 0.7 m_\mu$,
	respectively.
	}
	\label{fig:muMDM_couplings}
\end{figure}

Recently the Fermilab muon $g-2$ experiment confirmed the Brookhaven result.
The discrepancy between the experimental measurements and the SM prediction of the muon
 magnetic moment $a_\mu$ is now given by~\cite{Davier:2017zfy,Colangelo:2018mtw,Hoferichter:2019mqg,Davier:2019can,Keshavarzi:2019abf,
Aoyama:2020ynm,Muong-2:2021ojo}\footnote{
	The recent lattice results indicate that there is a discrepancy in the hadronic vacuum polarization
	contribution determined by the data-driven method and the lattice QCD
	method~\cite{Borsanyi:2020mff,Ce:2022kxy,Alexandrou:2022amy,FermilabLattice:2022smb}.
	In this paper, we assume that 
	the data-driven estimation~\eqref{eq:muonMDM_discrepancy} is correct.
}
\begin{align}
	\Delta a_\mu = a_\mu^\text{exp} - a_\mu^\text{SM} = \brr{251 \pm 59} \times 10^{-11}.
	\label{eq:muonMDM_discrepancy}
\end{align}
In our model, $\phi$ contributes to the anomalous muon magnetic dipole moment at one-loop as
\begin{align}
	a_\mu^{(\phi)} = \frac{m_\mu}{8\pi^2} \int_0^1 dz
	\frac{(1-z)^2\srr{(\abs{g_V}^2 + \abs{g_A}^2)z m_\mu + (\abs{g_V}^2 - \abs{g_A}^2)m_l }}
	{-z(1-z) m_\mu^2 + (1-z) m_l^2 + z m_\phi^2}.
\end{align}
This contribution can explain the discrepancy~\eqref{eq:muonMDM_discrepancy}
for both $l = e$ and $\tau$ cases, as shown in Fig~\ref{fig:muMDM_couplings}.
The above formula agrees with~\cite{Endo:2020mev} 
in the limit $m_l \ll m_\mu, m_\phi$.

In our model, $\phi$ induces other electromagnetic dipole moments as well.
In particular, $\phi$ induces the anomalous magnetic moment of $l$ as
\begin{align}
	a^{(\phi)}_l &=  \frac{m_l}{8\pi^2}\int_0^1 dz
	\frac{(1-z)^2\left[\left(\abs{g_V}^2+\abs{g_A}^2\right)z m_l
	+ \left(\abs{g_V}^2-\abs{g_A}^2\right)m_\mu\right]}
	{-z(1-z)m_l^2  + (1-z) m_\mu^2+zm_\phi^2}.
	\label{eq:al}
\end{align}
The anomalous magnetic dipole moment of $\tau$ is only limited as 
$-0.052 < a_\tau < 0.013$ by LEP~\cite{DELPHI:2003nah} and thus we safely ignore the contribution of $\phi$ to $a_\tau$.
In contrast, the anomalous magnetic dipole moment of $e$ is measured 
with extreme precision~\cite{VanDyck:1987ay,Odom:2006zz,Hanneke:2008tm,Fan:2022eto}.
Given the discrepancy between different measurements of the fine structure constant~\cite{Parker:2018vye,Morel:2020dww}
that is crucial for the SM prediction of $a_e$,
we may require that the contribution from $\phi$ satisfies
\begin{align}
	-3.4\times 10^{-13} < a_e^{(\phi)} < 9.8\times 10^{-13},
	\label{eq:eMDM}
\end{align}
following~\cite{Morel:2020dww}.
Since $a_e^{(\phi)}$ picks up the chirality flip from the internal muon unless the coupling is 
purely left-handed or right-handed, corresponding to the second term in the numerator of Eq.~\eqref{eq:al},
the constraint on $a_e^{(\phi)}$ prefers $\vert{g_V}\vert \sim \vert g_A \vert$,
as we show in Fig.~\ref{fig:muMDM_couplings}.
Therefore we focus on this parameter region in the $\mu\hyphen e$ case.
Moreover, $\phi$ induces the electric dipole moments (EDMs) of the muon and $l$ as
\begin{align}
	d_\mu
	&= -\frac{\mathrm{Im}(g_V^*g_A)m_l}{8\pi^2}\int_0^1dz
	\frac{(1-z)^2}{-z(1-z)m_\mu^2 + (1-z)m_l^2 + zm_\phi^2},
	\\
	d_l
	&= -\frac{\mathrm{Im}(g_V^*g_A)m_\mu}{8\pi^2}\int_0^1dz
	\frac{(1-z)^2}{-z(1-z)m_l^2 + (1-z)m_\mu^2 + zm_\phi^2}.
\end{align}
The electron EDM is constrained with high precision~\cite{ACME:2018yjb}, while the muon and tau EDM
can be constrained both directly and indirectly~\cite{Belle:2002nla,Muong-2:2008ebm,Ema:2021jds,Ema:2022wxd}.
For simplicity, we assume no CP violation, $\mathrm{Im}[g_V^* g_A] = 0$,
and ignore the EDMs in the following.
Then we take both $g_V$ and $g_A$ real by redefining the phase of $\phi$
without loss of generality.

The result of this subsection sets the sizes of the couplings of interest.
In the following, 
we will see that the whole parameter region is excluded by muonium invisible decay
and the LSND experiment in the $\mu\hyphen e$ case.
Although previous and current experiments do not exclude the $\tau\hyphen \mu$ case,
the whole parameter space can be explored by future experiments such as SHiP.

\section{$\mu\hyphen e$ case: Muonium and LSND}
\label{sec:mu-e}
In this section we consider the case that $\phi$ couples to $\mu$ and $e$, \emph{i.e.} 
the $\mu\hyphen e$ case.
We will see that the muonium invisible decay branching ratio puts a constraint on the couplings.
More interestingly, muonium formation and its subsequent decay produce 
a sizable flux of $\phi$ at neutrino experiments with muon decay at rest,
such as the Liquid Scintillator Neutrino Detector (LSND) experiment. 
Even though $\phi$ has a huge decay length, as we see below,
a small portion of $\phi$-decay inside the LSND detector is enough to exclude the entire parameter region
that explains the muon $g-2$ anomaly.

\subsection{Decay rates}
In this subsection, 
we compute the decay rates of $\phi$ and muonium into $\phi$  that are necessary to understand the phenomenology of $\phi$.

\subsubsection{$\phi$ decay rate}

\begin{figure}[t]
\centering
\includegraphics[width=0.5\linewidth]{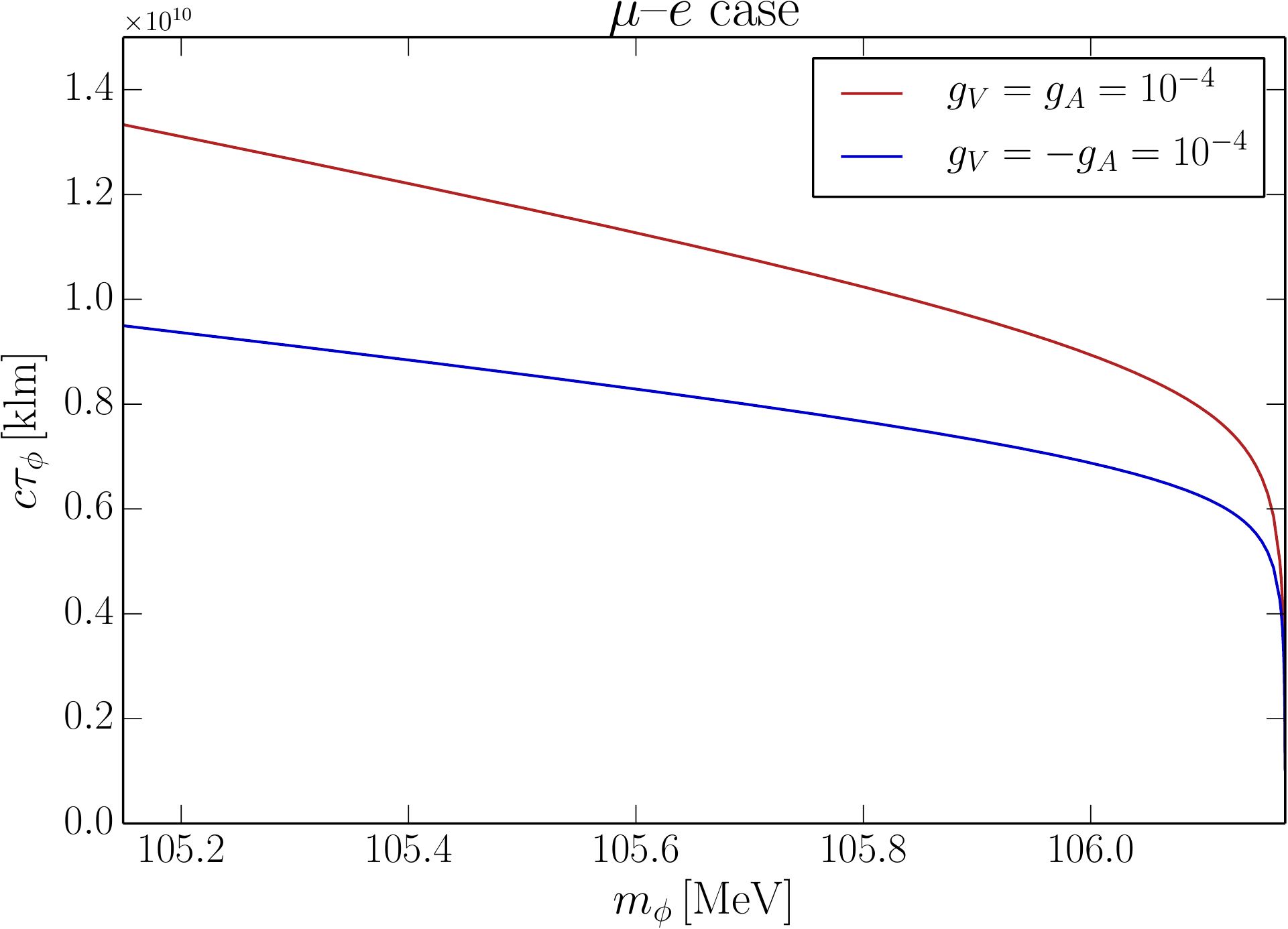}
\caption{The decay length $c \tau_\phi$ of $\phi$ in the $\mu \hyphen e$ case.
The red line corresponds to $g_V = g_A = 10^{-4}$
and the blue line to $g_V = -g_A = 10^{-4}$.
Note that the parameter region $\abs{g_V} \simeq \abs{g_A}$ is preferred from the electron $g-2$ constraint.}
\label{fig:mu-e-scalar-decay}
\end{figure}

Since we consider the mass range $m_\mu - m_e < m_\phi < m_\mu + m_e$,
the two-body decay $\phi \to \mu^- e^+$ is not allowed, and thus $\phi$ becomes accidentally long-lived.
Instead, $\phi$ mainly decays as $\phi \to e^- e^+ \nu_\mu \bar{\nu}_e$ through an off-shell $\mu^-$,
whose amplitude is diagrammatically given by
\begin{align}
	i\mathcal{M} = \begin{tikzpicture}[baseline=(v1)]
	\begin{feynman}[inline = (base.v1)]
		\vertex [label=\({\scriptstyle \phi}\)] (phi);
		\vertex [right = of phi] (v1);
		\vertex [above right = of v1, label=360:\({\scriptstyle e^+}\)] (ep);
		\vertex [below right = of v1] (v2);
		\vertex [above right = of v2, label=360:\({\scriptstyle e^-}\)] (em);
		\vertex [right = of v2, label=360:\({\scriptstyle \nu_\mu}\)] (numu);
		\vertex [below right = of v2, label=360:\({\scriptstyle \bar{\nu}_e}\)] (nue);
		\node [left = -0.05cm of v2, dot,] (fermi);
		\begin{pgfonlayer}{bg}
		\diagram*{
		(phi) -- [charged scalar] (v1) -- (ep),
		(v1) -- [fermion] (v2) -- [fermion] (numu),
		(v2) --(em), (v2) -- (nue),
		};
		\end{pgfonlayer}
	\end{feynman}
	\end{tikzpicture},
\end{align}
where the dot indicates the four-Fermi operator and the arrow indicates the flow of the muon number.
The decay width of $\phi$ is given by (see App.~\ref{app:phi_decay} for derivation)
\begin{align}
	\Gamma_\phi &= \frac{G_F^2}{384\pi^5 m_\phi}
	 \int_{m_e^2}^{(m_\phi-m_e)^2}\frac{d\bar{m}_\mu^2}{(\bar{m}_\mu^2 - m_\mu^2)^2}
	\int_{0}^{(\bar{m}_\mu-m_e)^2}d\bar{m}_{\nu\nu}^2 I_{1}(\bar{m}_\mu, \bar{m}_{\nu\nu})
	\nonumber \\
	&\times
	\left[\left(\abs{g_V+g_A}^2 \bar{m}_\mu^2 + \abs{g_V-g_A}^2 m_\mu^2\right) I_{2}(\bar{m}_\mu)
	-2(\abs{g_V}^2 - \abs{g_A}^2)m_e m_\mu \bar{m}_\mu^2 I_{3}(\bar{m}_\mu)\right],
	\label{eq:Gamma_phi_mue}
\end{align}
where
\begin{align}
	I_{1} &= \left(1+\frac{\bar{m}_{\nu\nu}^2-2m_e^2}{\bar{m}_\mu^2}
	+ \frac{m_e^4 + m_e^2 \bar{m}_{\nu\nu}^2 - 2\bar{m}_{\nu\nu}^4}{\bar{m}_\mu^4}
	\right)
	\sqrt{\left(\bar{m}_\mu^2 - \bar{m}_{\nu\nu}^2\right)^2
	- 2m_e^2 (\bar{m}_\mu^2 + \bar{m}_{\nu\nu}^2) + m_e^4},
	\\
	I_{2} &= \frac{m_\phi^2 - \bar{m}_\mu^2 - m_e^2}{4m_\phi^2}\sqrt{
	(m_\phi^2 - \bar{m}_\mu^2)^2 - 2 m_e^2 (m_\phi^2 + \bar{m}_\mu^2) + m_e^4
	},
	\\
	I_{3}
	&= \frac{\sqrt{(m_\phi^2 - \bar{m}_\mu^2)^2 - 2 m_e^2 (m_\phi^2 + \bar{m}_\mu^2) + m_e^4}}
	{2m_\phi^2}.
\end{align}
In this formula $G_F$ is the Fermi constant,
$\bar{m}_\mu$ is the invariant mass of the off-shell muon, and $\bar{m}_{\nu\nu}$
is the invariant mass of the $\nu_\mu \bar{\nu}_e$ system, respectively.
We plot the decay length of $\phi$ in Fig.~\ref{fig:mu-e-scalar-decay}.
As one can see, $\phi$ has a huge decay length compared to the detector length due to the four-body decay phase space and the smallness of the couplings. 

\subsubsection{Muonium decay rate}
\label{subsec:muonium_decay}

Muonium is a hydrogen-like bound state consisting of $\mu^+$ and $e^-$.
The mass range $m_\mu - m_e < m_\phi < m_\mu + m_e$ (ignoring the binding energy) allows muonium to decay into $\phi^*$ and photon.
The muonium decay rate into $\phi^* \gamma$ can be written as
\begin{align}
	\Gamma(\mathrm{Mu}\to \phi^* \gamma) 
	&= \abs{\psi(0)}^2 \times \sigma_\mathrm{free} v(\mu^+ e^- \to \phi^* \gamma),
\end{align}
where $\sigma_\mathrm{free}(\mu^+ e^- \to \phi^* \gamma)$ is the annihilation cross-section of free $\mu^+$ and $e^-$ going to $\phi^*$ and $\gamma$,  and $\psi(0)$ is the muonium wavefunction at the origin, given by
\begin{align}
	\abs{\psi(0)}^2 &= \frac{\alpha^3 m_e^3}{\pi},
\end{align}
for the ground state in the limit $m_e \ll m_\mu$.
In the non-relativistic limit, we obtain (see App.~\ref{app:muonium_decay} for derivation)
\begin{align}
	\Gamma\left(\mathrm{Mu}^{(S=1)}\to \phi^* \gamma\right) &=
	\frac{\alpha^4(\abs{g_V}^2 + \abs{g_A}^2)}{3\pi}\frac{m_e (m_\mu + m_e - m_\phi)}{m_\mu},
	\label{eq:muonium_decay1} 
	\\
	\Gamma\left(\mathrm{Mu}^{(S=0)}\to \phi^* \gamma\right) &= 0,
	\label{eq:muonium_decay2}
\end{align}
where the superscripts $S = 0$ and $1$ indicate the spin of the muonium.
Here the selection rule works due to the angular momentum conservation,
in a similar way as the positronium decay into photons.
The muonium lifetime is dominated by the individual muon lifetime, 
and thus, the branching ratio is given by
\begin{align}
	\mathrm{Br}(\mathrm{Mu}^{(S=1)} \to \phi^* \gamma) 
	&= 2.5\times 10^{-5}\times 
	\frac{m_\mu + m_e - m_\phi}{m_e}\left(\frac{\abs{g_V}^2 + \abs{g_A}^2}{10^{-8}}\right).
\end{align}
This rate is sizable since the decay into $\phi^*$ competes with the weak process, 
which results in $G_F^2$ in the denominator.

\subsection{Constraints}

We now discuss the constraints of our model.
We will see that the parameter region that explains the muon $g-2$ anomaly is excluded by muonium invisible decay and by the LSND experiment.

\subsubsection{Muonium invisible decay}

As we saw above, muonium can decay into $\phi^*$ and $\gamma$, 
even though a free $\mu^+$ cannot decay into $\phi^* e^+ \gamma$.
Ref.~\cite{Gninenko:2012nt} sets a bound on such an invisible 
(\emph{i.e.} $e^+$--less) decay branching ratio of muonium based on the results of the MuLan experiment~\cite{MuLan:2007qkz,MuLan:2010shf,MuLan:2012sih}.
The MuLan experiment performed a dedicated study on the muon lifetime with two different stopping materials, the magnetized ferromagnetic alloy target that does not host muonium formation,
 and the quartz inside which a stopped $\mu^+$ forms muonium.
The experiment did not find any difference in the lifetime between the two materials,
from which~\cite{Gninenko:2012nt} derived an upper bound as
\begin{align}
	\mathrm{Br}(\mathrm{Mu}^{(S=1)}\to \mathrm{invisible}) &< 5.7\times 10^{-6}.
	\label{eq:muonium_invisible}
\end{align}
This constraint directly applies to our model.

As we see now, the muonium decay into $\phi^* \gamma$ 
has another interesting implication. 
While the produced $\phi$ is simply invisible in the muonium decay experiments, 
it leaves interesting visible signatures at the LSND neutrino experiment.

\subsubsection{LSND}
The LSND experiment
is designed to search for $\bar{\nu}_\mu \rightarrow \bar{\nu}_e$ oscillation~\cite{LSND:1996jxj}.
Neutrinos are produced from the decay of stopped $\pi^+$ and $\mu^+$. 
In particular, $\bar{\nu}_\mu$ is primarily from the decay of stopped $\mu^+$ which originates from the stopping $\pi^+$.
The detector is cylindrical in shape, with a diameter $5.7\,\mathrm{m}$ 
(\emph{i.e.} the area $A_\mathrm{LSND} = \pi (5.7/2)^2\,\mathrm{m}^2$) 
and length $l_\mathrm{LSND} = 8.3\,\mathrm{m}$~\cite{LSND:2001akn}.
It is located 29.8~m downstream of the proton beam stop at an angle of $12^\circ$ to the proton beam. 
In our model, a non-relativistic flux of $\phi^*$ is produced 
from the muonium decay once $\mu^+$ gets trapped and forms muonium at the beam stop.\footnote{
	The pion decay also produces a flux of $\phi$, but this is far more suppressed as the branching ratio of $\pi$ decaying into $\phi$ is small.
	This is because one has to pay $G_F$ for this decay mode, and hence there is no $1/G_F^2$ enhancement of the branching ratio that we had in the case of the muonium decay.
}
Once produced, a small portion of $\phi^*$ flies to the downstream detector and decays
inside the detector, leaving $e^+$ signals that mimic the $\nu_e \hyphen e$ elastic scattering.\footnote{
	The decay of $\phi^*$ produces $e^+ e^-$, 
	but $e^-$ is soft since the decay prefers almost on-shell intermediate $\mu^+$.
} 

The total number of $\nu_e$ that passes through the LSND detector is given by~\cite{LSND:2001akn}
\begin{align}
	N_{\nu_e} = \Phi_{\nu_e} \times A_{\mathrm{LSND}},
	\quad
	\Phi_{\nu_e} = 1.176\times 10^{14}\,\mathrm{cm}^{-2}.
\end{align}
Since $\nu_e$ dominantly comes from $\mu^+$,
we can rescale $N_{\nu_e}$ by the muonium formation probability and the branching ratio
to estimate the total number of $\phi^*$ that pass through the LSND detector as
\begin{align}
	N_{\phi^*} &= N_{\nu_e} P\left(\mu^+ \to \mathrm{Mu}^{(S=1)}\right)\mathrm{Br}\left(\mathrm{Mu}^{(S=1)}\to \phi^* \gamma\right),
\end{align}
where $P(\mu^+ \to \mathrm{Mu}^{(S=1)})$ is the triplet muonium formation probability 
which depends on the stopping material.
Note that $\Phi_{\nu_e}$ already takes into account the angular coverage of the detector.
The differential event number is estimated as
\begin{align}
	\frac{d^2 N_e}{d E_e d\cos\theta_e}
	&= N_{\phi^*}\frac{l_\mathrm{LSND}}{\lambda_\phi} 
	\frac{1}{\Gamma_\phi}\frac{d^2 \Gamma_\phi}{d E_e d\cos\theta_e}.
\end{align}
Here the mean decay length $\lambda_\phi$ is defined as
\begin{align}
	\lambda_\phi = \gamma \beta c \tau_\phi
	\simeq \frac{m_\mu + m_e - m_\phi}{m_\phi}\frac{1}{\Gamma_\phi},
\end{align}
where we take the limit $m_e \ll m_\mu$ and ignore the binding energy of muonium.
Note that the produced $\phi$ is non-relativistic with velocity $\beta \sim \mathcal{O}(m_e/m_\mu)$,
which enhances the probability of $\phi$ decaying inside the decay volume by $1/\beta$.

\begin{figure}
\centering
\includegraphics[width=0.475\linewidth]{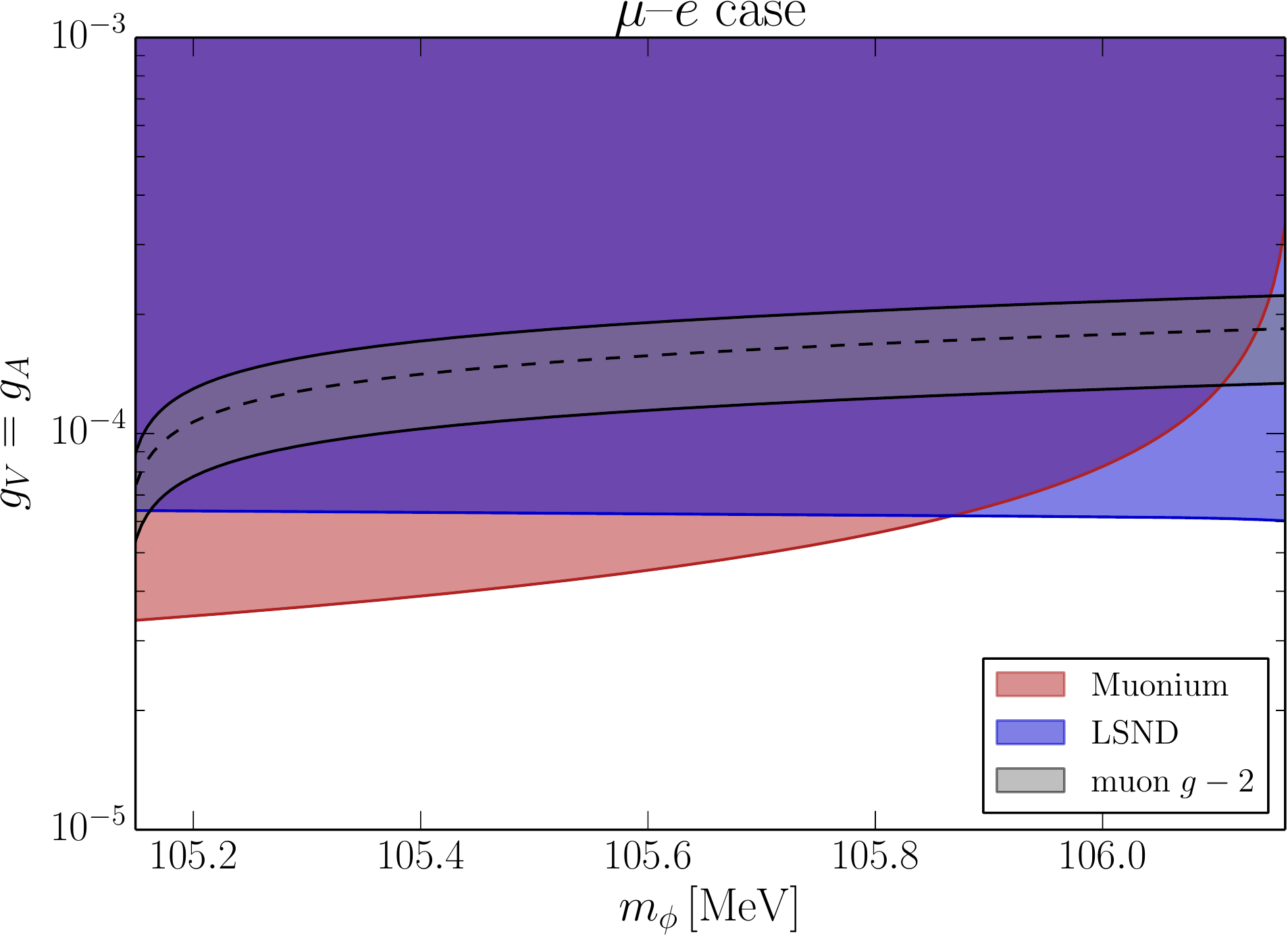}
\hspace{0mm}
\includegraphics[width=0.475\linewidth]{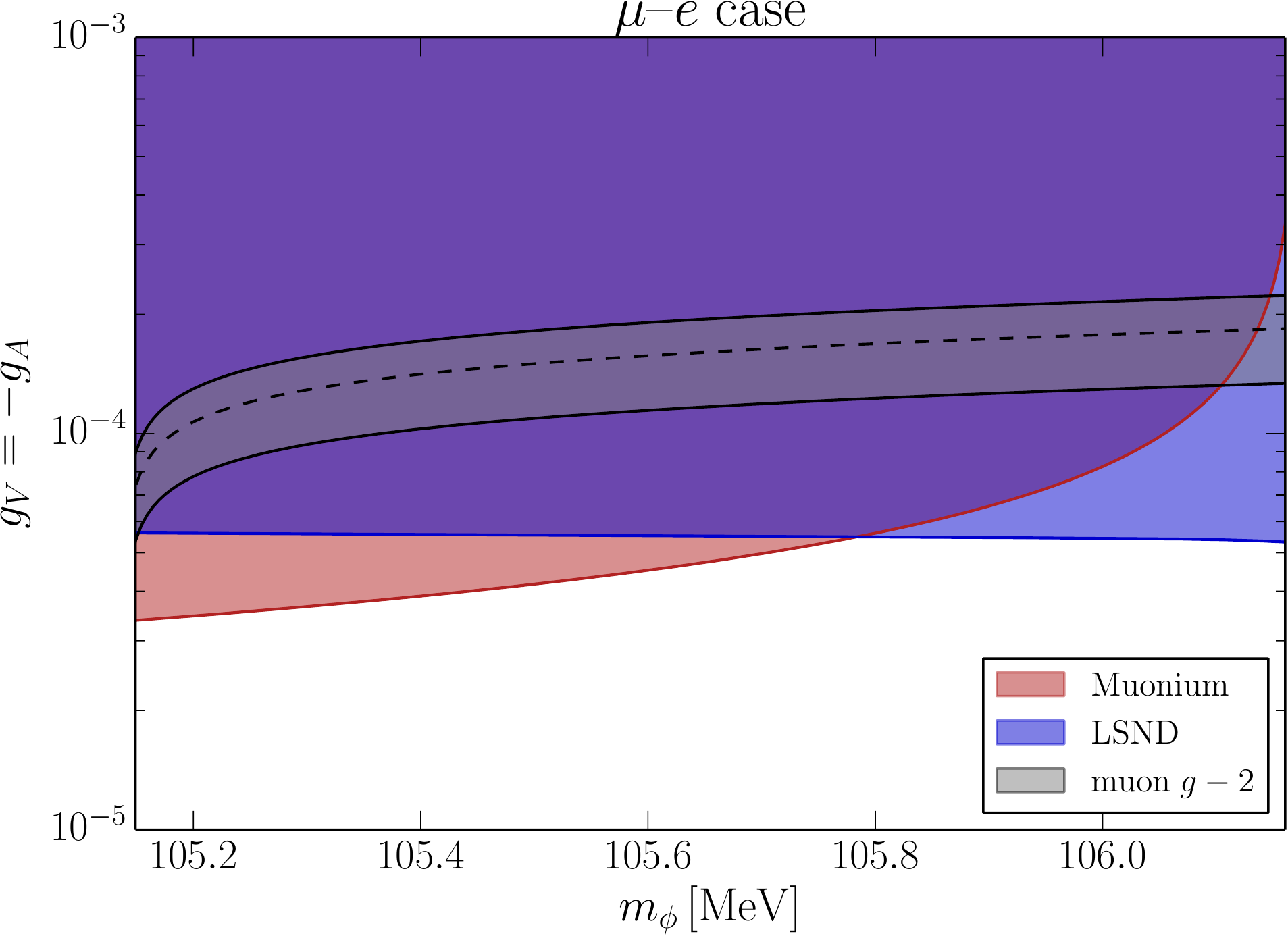}
	\caption{
	The constraints on the couplings in the $\mu\hyphen e$ case.
	The region excluded by the muonium invisible decay and the LSND
	is covered in red and blue, respectively.
	The black dashed line corresponds to the central value of the muon $g-2$ anomaly,
	and the gray region indicates the $1\sigma$ band.
	The left panel is with $g_V = g_A$ while the right panel is with $g_V = - g_A$.
	Note that only the LSND constraint is (mildly) affected by the relative sign between $g_V$ and $g_A$.
	One can see that the whole parameter region favored by the muon $g-2$ is excluded.}
\label{fig:LSND_mue}
\end{figure}

The decay of $\phi$ in the detector volume will create a very soft electron and a positron with energies closely following the Michel positron spectrum for the on-shell muon decay. Thus $\phi$ decay will appear as a single energetic $e^+$ in that detector. Therefore, one should look for a signature of $\phi$ decays in the sample of neutrino-electron scattering events, $\nu e \to \nu e$, as electron and positron events are nearly identical.
The observed spectrum of beam excess events, consistent with the neutrino scattering on electrons at the LSND experiment, is reported in~\cite{LSND:2001akn}.
In particular, the experiment observed 49 events 
in the energy bin $20\,\mathrm{MeV} < E_e < 22\,\mathrm{MeV}$
with $\cos\theta_e > 0.9$.
Therefore, we can derive a conservative constraint on the coupling, without subtracting $\nu e$ scattering and by requiring that
\begin{align}
	N_e &= \int_{20\,\mathrm{MeV}}^{22\,\mathrm{MeV}} dE_e \int_{0.9}^{1}d\cos\theta_e
	\frac{d^2 N_e}{d E_e d\cos\theta_e}
	< 49.
	\label{eq:LSND}
\end{align}
If we include the other energy bins and energy spectrum information, the constraint will get stronger,
but the above condition is enough for our purpose, to exclude the region relevant for the muon $g-2$.
In Fig.~\ref{fig:LSND_mue}, 
we plot the constraint from the muonium invisible decay~\eqref{eq:muonium_invisible}
and the LSND experiment~\eqref{eq:LSND},
where we take $P(\mu^+ \to \mathrm{Mu}^{(S=1)}) = 3/4$~\cite{Gninenko:2012nt}.
As one can see, the whole parameter region that explains the muon $g-2$ is excluded.
Our constraint relies on two distinct observables and hence is robust.
We have checked that the parameter region $\abs{g_V} \gtrsim \abs{g_A}$
that can explain muon $g-2$ without spoiling the electron $g-2$ measurement
is also excluded by the muonium invisible decay and the LSND experiment.

\section{$\tau\hyphen\mu$ case: Beam dump experiments}
\label{sec:tau-mu}
We now move to the $\tau\hyphen\mu$ case, for which the considerations are much different from the $e\hyphen\mu$ case.
In this case, high energy beam dump experiments such as CHARM, NuTeV, and SHiP that can produce a sizable amount of $\tau$ can also produce $\phi$,
and its subsequent decay at downstream detectors can be observed and used to probe the model.
As we will see, although the CHARM and NuTeV experiments do not exclude the parameter region interesting for the muon $g-2$ anomaly, the future SHiP experiment can cover the whole parameter space.

\begin{table}[t]
  \centering
  \caption{Major experiments and parameters for the $\tau-\mu$ case, see details in the text. For comparison, we list the geometric acceptance for the $D_s$ production at CHARM and SHiP experiments.
  }\vspace{0.2cm}
  \resizebox{\textwidth}{!}{
    \begin{tabular}{|c|c|c|c|c|c|c|c|}
    \hline
    Experiment & \multicolumn{1}{l|}{$E_{\rm beam}$(GeV)} & POT   & \multicolumn{1}{l|}{$D$ (m)} & \multicolumn{1}{l|}{$L$ (m)} & $A$ ($m^2$) & $\epsilon^{\rm geo}_{\rm acpt}$ & Major production \\
    \hline
    CHARM & 400   & $2\times10^{18}$ & 480   & 35    &   3$\times$3    & 1.3\% & EW, $D_s$, $\mu$OT \\
    \hline
    NuTeV & 800   & $\mu$OT & 850   &  34     & $2.54 \times 2.54 $     &  $\mathcal{O}$(1)\%     & $\mu$OT \\
    \hline
    SHiP  & 400   & $2\times10^{20}$ & 60    & 60    & 5$\times$10  & 54\% & EW, $D_s$, $\mu$OT \\
    \hline
    \end{tabular}%
    }
  \label{tab:taumuproduction}%
\end{table}%

\subsection{Production mechanism}

First, we discuss the production of $\phi$ at beam dump experiments.
We focus on high-energy beam dump experiments such as the past CHARM and NuTeV experiments
as well as the future SHiP experiment, as high energy is required to produce $\phi$ 
with its mass $m_\phi \sim m_\tau$.
In these experiments, we notice three ways of producing $\phi$: 
the direct electroweak process, heavy meson ($D_s$ dominant) decay, and high energy muon passing through dense matter, 
which we may refer to as ``muon on target." 

\begin{figure}[t]
    \centering
    \includegraphics[width=\linewidth]{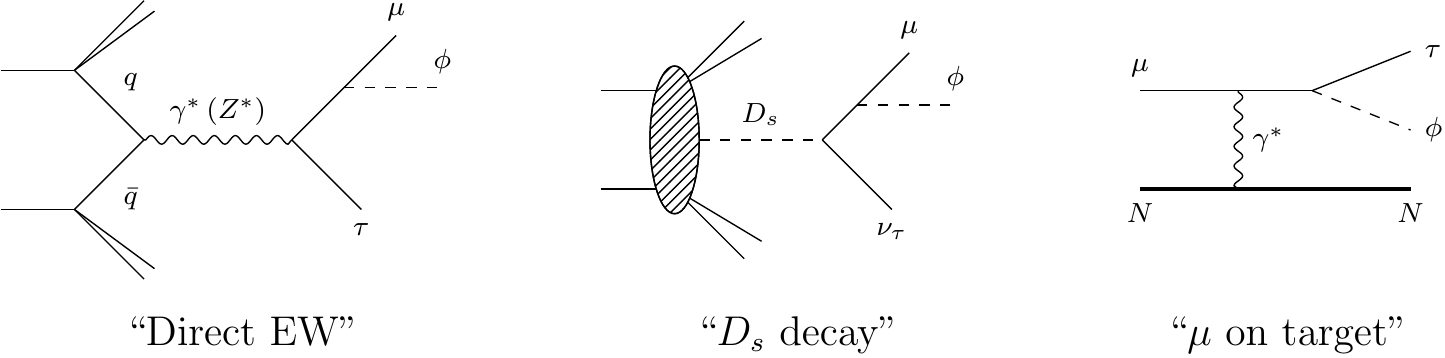}
    \caption{Primary production modes and representative Feynman diagrams for each mode for the $\mu$-$\tau$-flavored scalar $\phi$ at beam dumps.}
    \label{fig:feynmandiagrams}
\end{figure}

We show schematically representative diagrams for each production mode in \autoref{fig:feynmandiagrams}. Depending on the detailed beam and detector setup, different mode dominates the detectable $\phi$ fluxes, which we summarize in \autoref{tab:taumuproduction}. We explain each process in this section.

\subsubsection{Direct electroweak production}

With a high energy beam, the direct weak interaction production from parton scatterings is one of the primary sources for the $\phi$ flux.
The electroweak process is with initial states quark pair colliding to $\phi$ plus lepton pairs, namely, $q \bar{q} \rightarrow \tau^\mp \mu^\pm \phi^{(*)}$. 
In the $\sqrt{s}$ energy regime suitable for the beam dumps considered in this study, the main contribution is through the $s$-channel off-shell photon diagram. 
Given the minimal partonic center of mass energy required here is above $\sim 2 m_\tau=3.4$~GeV, the parton model works well to describe the production cross section here. We choose {\tt MadGraph}~\cite{Alwall:2011uj} to simulate the direct weak production cross section and kinematics distributions. A typical Feynman diagram is shown in the left panel of \autoref{fig:feynmandiagrams}. 

\begin{figure}[t]
    \centering
    \includegraphics[width=8cm]{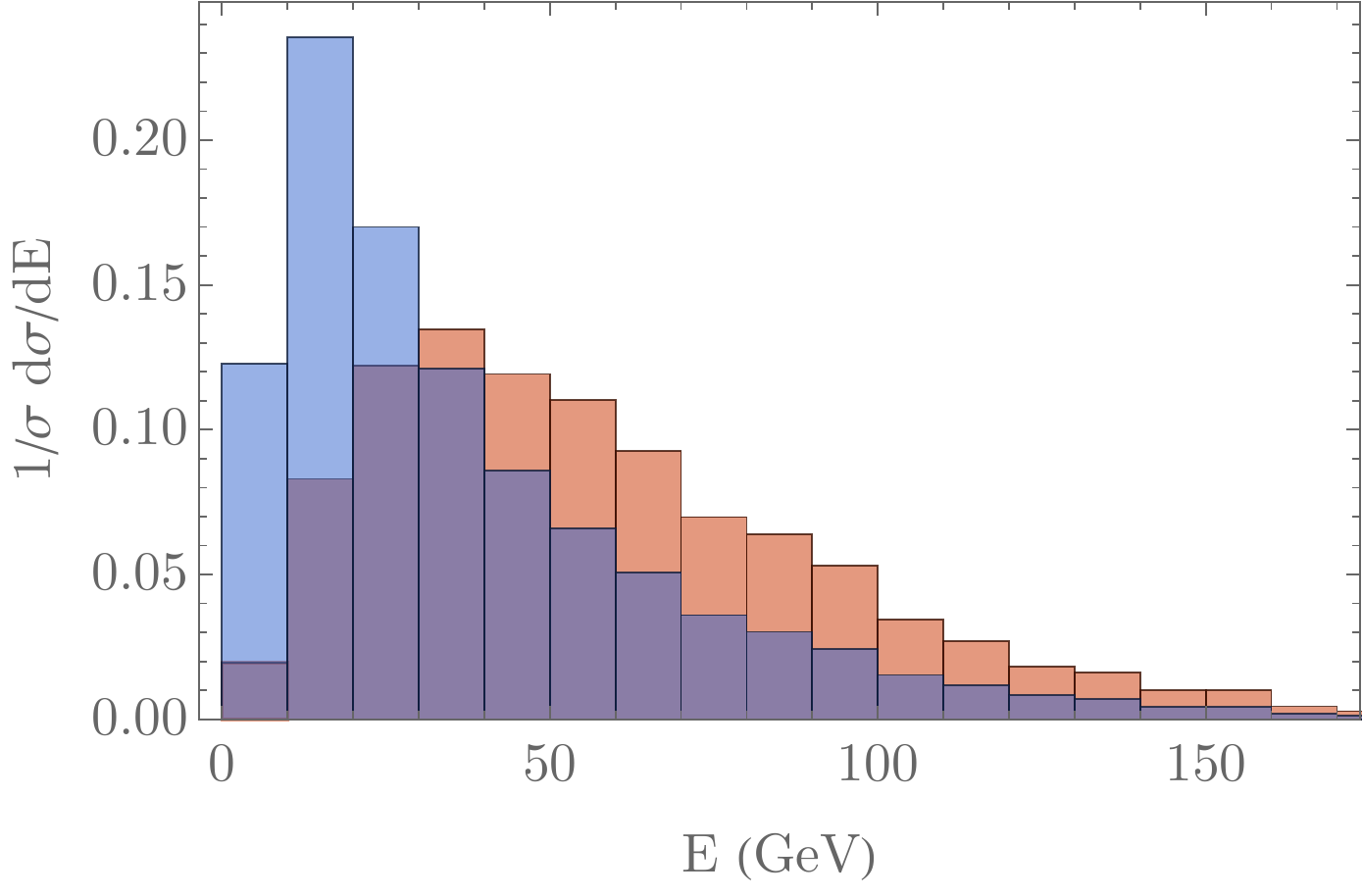}
    \caption{The normalized differential energy distribution from the direct EW production at SHiP experiment. The blue and red distribution are before and after the angular acceptance selection. The geometric acceptance slightly favors higher energy with a signal geometric acceptance efficiency of 42\% for SHiP.}
    \label{fig:EWdistribution}
\end{figure}

Given the center of mass energy of the beam-on-target collisions is around $15-20$~GeV, a partonic center of mass energy above 3.4~GeV corresponds to a sizable $x$ in the parton distribution functions. Hence, $\phi$ productions are typical with small transverse momentum, making the geometric acceptance of near detectors of various experiments sizable. To illustrate a typical energy distribution, we show in \autoref{fig:EWdistribution} the normalized energy distribution of the produced $\phi$, which has very little dependence on the $\phi$ mass in the allowed mass window. We show the distribution at production in blue and the geometrically accepted distribution by the SHiP experiment in red. We can see that geometric selection favors slightly harder $\phi$'s as there is a typical transverse momentum associated with the production. For the case of CHARM and NuTeV, the geometric selection will be further biased towards the high energy part. The production is off-shell photon exchange dominated, given that the massive gauge boson mediated processes (both neutral and charged current) are highly suppressed by $\hat s^2/m_{W, Z}^4$ (and $\hat s/m_{W, Z}^2$ for the interference). We hence do not include the charged current production in this study.

\subsubsection{Heavy meson decay}

\begin{figure}[t]
    \centering
    \includegraphics[width=0.5\linewidth]{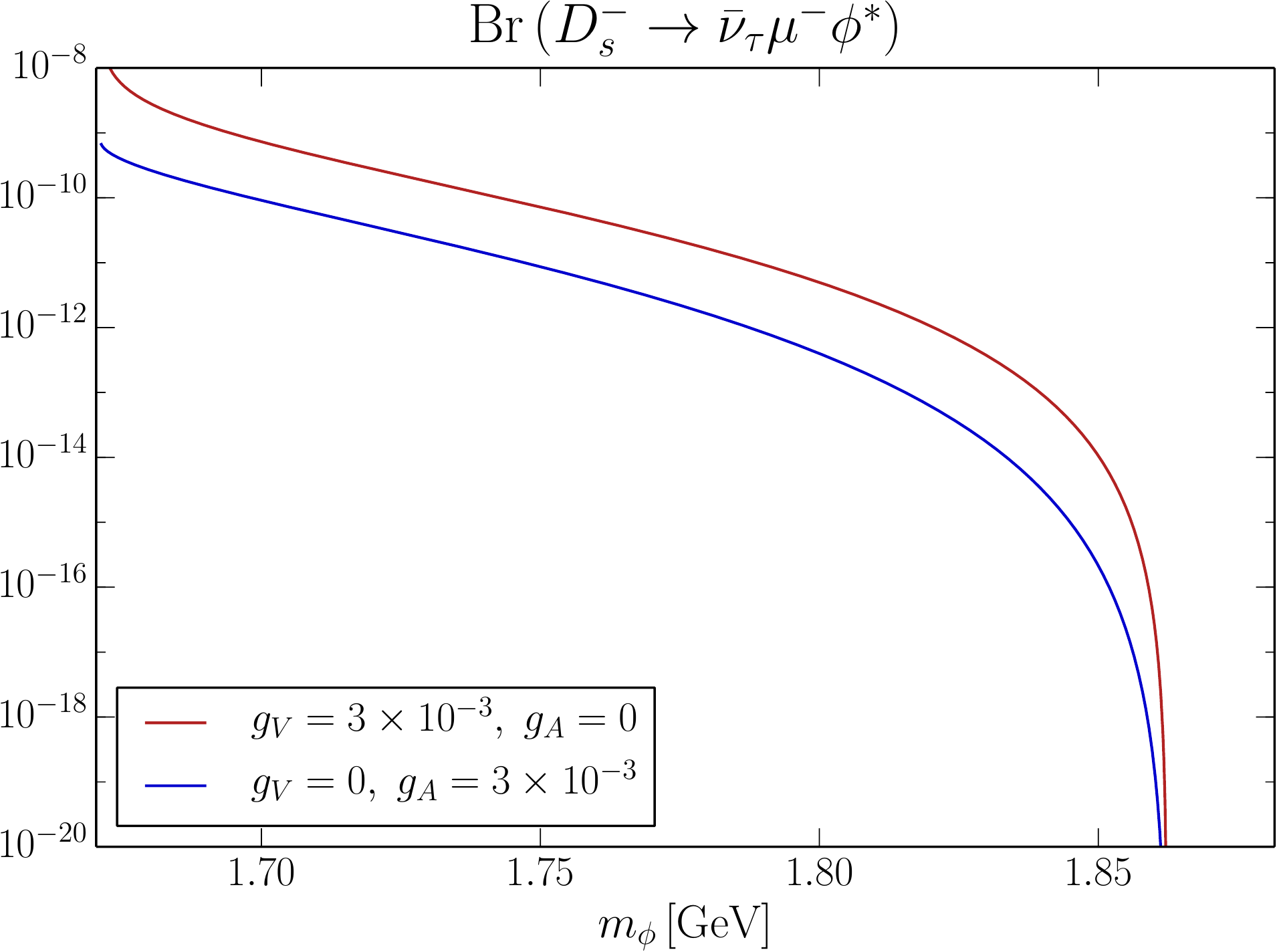}
    \caption{The branching ratio of $D_s$ decaying to $\nu_\tau \mu \phi$.
    The red line corresponds to $g_V = 3\times 10^{-3}$ and $g_A = 0$, while
    the blue line to $g_V = 0$ and $g_A = 3\times 10^{-3}$.
    Note that the branching ratio is strongly suppressed due to the phase space 
    in the large $m_\phi$ region, and eventually vanishes for $m_{D_s} - m_\mu < m_\phi < m_\tau + m_\mu$.}
    \label{fig:Ds_decay_br}
\end{figure}

In typical beam dump experiments, the primary source of $\tau$ is the decay of $D_s$ mesons,
since kaons and pions do not have enough mass while the decays of other $D$ meson
are suppressed by the CKM matrix element.
In a similar way, $D_s$ mesons can produce $\phi$ through off-shell $\tau$, 
whose amplitude is diagrammatically given by
\begin{align}
	i\mathcal{M}
	&= \begin{tikzpicture}[baseline=(a)]
	\begin{feynman}[inline = (base.a), horizontal=a to b]
		\vertex [label=\(\scriptstyle {D^-_s,\,P}\)] (a);
		\vertex [right = of a] (b);
		\vertex [below right = of b, label=360:\(\scriptstyle \bar{\nu}_\tau{,}\,p_3\)] (c);
		\vertex [above right = of b] (d);
		\vertex [below right = of d, label=360:\(\scriptstyle \mu^-{,}\,p_2\)] (e);
		\vertex [above right = of d, label=360:\(\scriptstyle \phi^*{,}\,p_1\)] (f);
		\diagram*{
		(a) -- [scalar] (b) -- (c),
		(b) -- (d) -- [fermion] (e),
		(f) -- [charged scalar] (d),
		};
	\end{feynman}
	\end{tikzpicture}.
\end{align}
The effective Lagrangian relevant for this process is induced from the charged current and is given by
\begin{align}
	\mathcal{L}_\mathrm{eff} &= -\frac{G_F}{\sqrt{2}}V_{cs}f_{D_s}
	(\partial_\mu D_s^-) \bar{\tau}\gamma^\mu(1-\gamma_5)\nu_{\tau}
	+ \mathrm{(h.c.)},
\end{align}
where $V_{cs} \simeq 0.95$ is the CKM matrix element and 
$f_{D_s} \simeq 255\,\mathrm{MeV}$ is the $D_s$ decay constant~\cite{Bernard:2000ki,ALEPH:2002fge}.
The amplitude is given by
\begin{align}
	i\mathcal{M}
	&= -\frac{G_F}{\sqrt{2}}V_{cs}f_{D_s}
	\bar{u}_\mu(p_2)\left(g_V+g_A\gamma_5\right)\frac{\slashed{p}_{12}+m_\tau}{m_{12}^2-m_\tau^2}
	\slashed{P}(1-\gamma_5)u_{\nu_\tau}(p_3),
\end{align}
where $p_{ij} = p_i + p_j$ and $m_{ij}^2 = p_{ij}^2$.
The decay width follows
\begin{align}
	\Gamma(D_s \to \phi \mu \nu_\tau)
	&= \frac{1}{256\pi^3 m_{D_s}^3} \int_{(m_\phi + m_\mu)^2}^{m_{D_s}^2}
	dm_{12}^2 \int_{(m_{23}^2)_+}^{(m_{23}^2)_-} dm_{23}^2 \abs{M}^2,
\end{align}
where
\begin{align}
	(m_{23}^2)_\pm &= \left(E_2^* + E_3^*\right)^2 - \left(\sqrt{E_2^{*2}-m_\mu^2} \pm E_3^*\right)^2,
\end{align}
with
\begin{align}
	E_2^* = \frac{1}{2m_{12}}\left(m_{12}^2 - m_\phi^2 + m_\mu^2\right),
	\quad
	E_3^* = \frac{1}{2m_{12}}\left(m_{D_s}^2 - m_{12}^2\right).
\end{align}
We plot the branching ratio in \autoref{fig:Ds_decay_br}.
The branching ratio strongly depends on $m_\phi$ due to the phase space suppression.
In particular, since $m_\tau + m_\mu > m_{D_s}-m_\mu$, 
the branching ratio vanishes in the large $m_\phi$ region.

\begin{figure}[t]
    \centering
    \includegraphics[width=7.5cm]{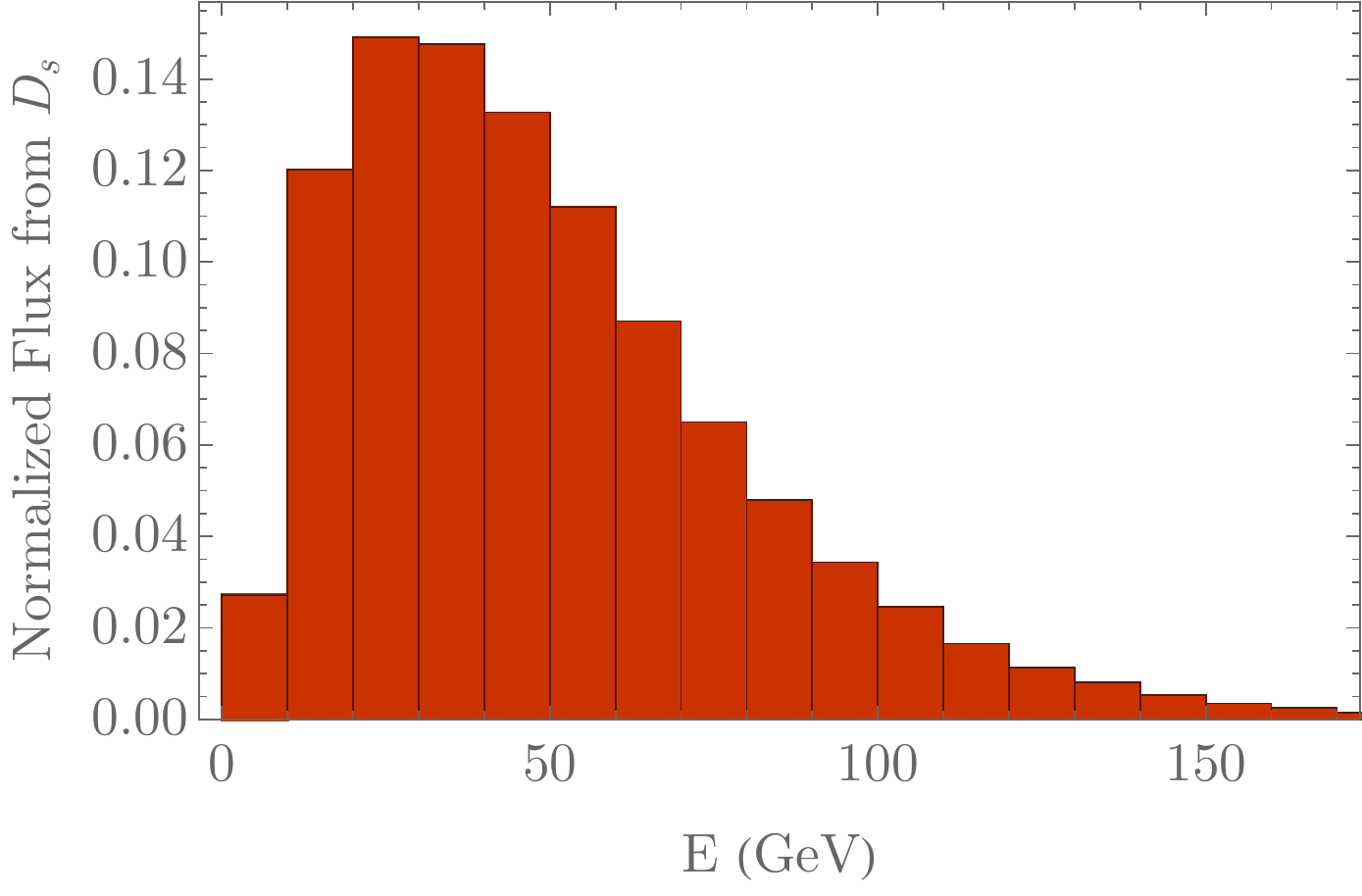}
    \includegraphics[width=7.5cm]{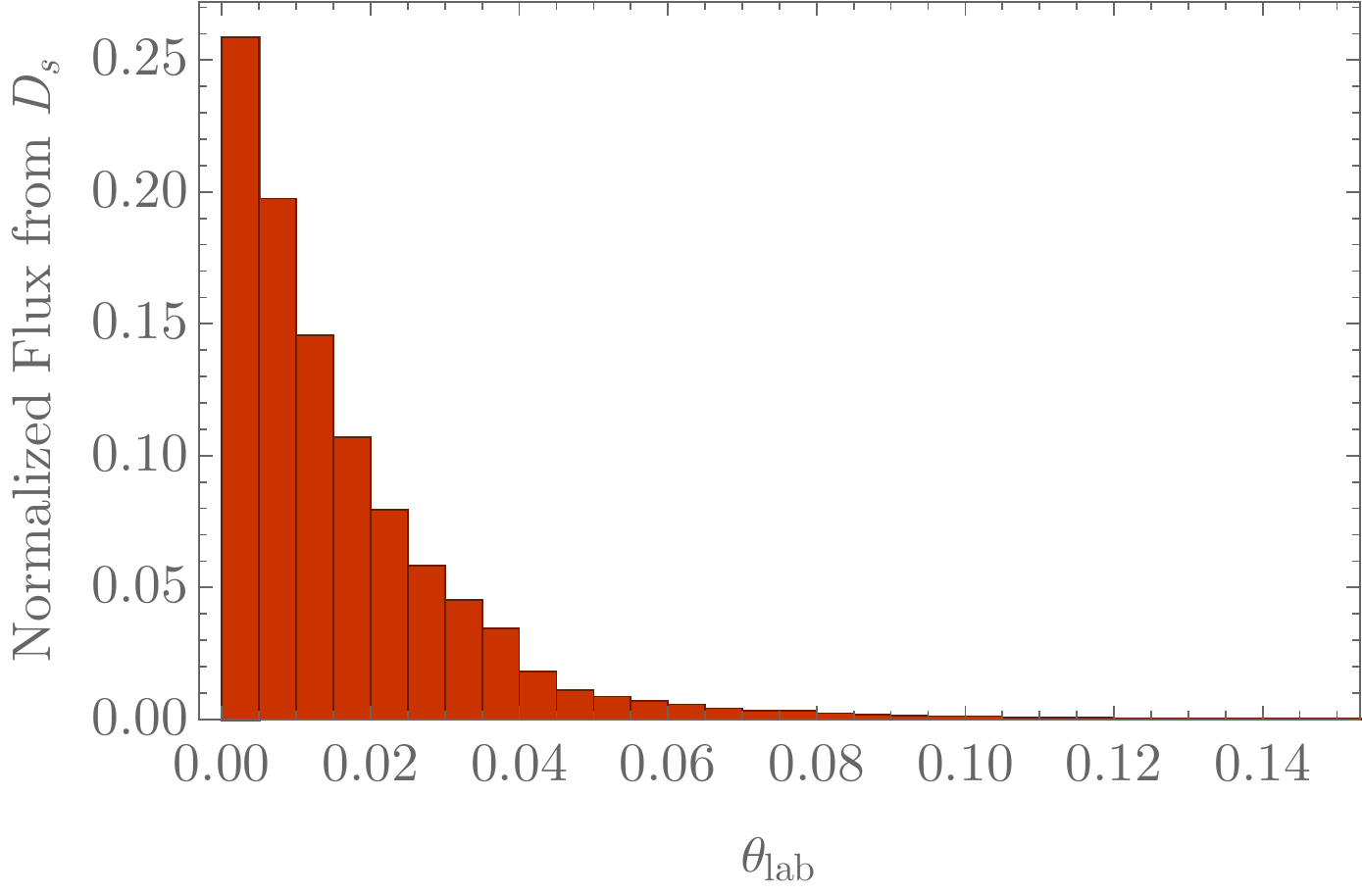}
    \caption{The normalized differential energy distribution and the angular distribution from the $D_s$ production at SHiP experiment.}
    \label{fig:Dsdistribution}
\end{figure}

In our numerical computation, we simulate the flux of $\phi$ from $D_s$ at the beam dump experiment using the combination {\tt Pythia8}~\cite{Sjostrand:2014zea,Bierlich:2022pfr} and our code. The $D_s$ production is driven by the soft-QCD process, which we use {\tt Pythia8} to simulate the cross section and kinematics. Then the number of $D_s$ produced at the target can be calculated as the product of the PoT, and we calculated the $D_s$ production cross section, and then divided by the total inclusive proton target-nucleus cross section of $\sigma_{pN}=10.7$~mb reported in Ref.~\cite{SHiP:2015vad}. We then decay the $D_s$ into $\phi+\tau+\mu$ in an approximation that ignores angular correlations (three-body phase space mapping) since we do not expect a strong angular correlation for the scalar decay. In \autoref{fig:Dsdistribution} we show the normalized differential energy distribution~(left panel) and angular distribution (right panel) at the SHiP experiment to show the typical behavior of $\phi$ from $D_s$ production. 
In the left panel, we can see that the typical energy of scalar produced through $D_s$ decay peaks around 30-40~GeV, with a long tail expanding to the hundreds of GeV regime. The spread of the energy distribution provides us access to both short- and long-lifetime regimes. The right panel shows that the $\phi$ production from $D_s$ decays does prefer the forward regions along the beam direction, enabling significant effective angular acceptance for the signal.
To verify and cross-check our numerical computation, we estimated the $D_s$ spectrum based on
the differential cross section of $pN \to c \bar{c} X$ and the $D_s$ formation rate provided in~\cite{Alekhin:2015byh},
assuming for simplicity that $c$ and $D_s$ have the same kinetic energy.
We have checked that the constraint derived from this procedure agrees well with our numerical computation.

\subsubsection{Muon on target}

In beam dump experiments, high-energy muon flux is produced mainly from the decay of mesons,
in particular from charged pions and kaons.
Once these high-energy muons pass through dense matters, the muons can produce $\phi$
by picking up virtual photons from nuclei,
corresponding to the process $\mu N \to \tau \phi N$. 
We phrase this production mechanism ``muon on target," or $\mu$OT.
In the NuTeV experiment, in order to shield high energy muons, there is 850\,m of dirt 
between the meson decay region and the detector,
and $\phi$ can be produced from muons passing through this dirt region.
In the SHiP experiment, there is $\sim$ 5\,m of steel as a beam stop after the target, 
before the active muon shield by a magnetic field, where $\phi$ can be produced.

The flux of $\phi$ at the detector produced by the $\mu$OT is estimated as
\begin{align}
	\Phi_\phi(E_\phi) &= \int dE\,\Phi_\mu (E) \int_0^{l_\mathrm{max}} dl\,n_A 
	\int_{0}^{\theta_\mathrm{det}(l)}d\theta_\phi \sin\theta_\phi 
	\frac{d^2\sigma}{dE_\phi d\cos\theta_\phi} (E_l, E_\phi),
\end{align}
where $\Phi_\mu(E)$ is the initial muon flux, 
$l_\mathrm{max} = \mathrm{min}[l_\mathrm{shield}, l_\mathrm{stop}(E)]$ with $l_\mathrm{shield}$ the length of the shield and $l_\mathrm{stop}(E)$
the length at which the shield stops a muon with initial energy $E$,
$n_A$ is the number density of the target nuclei, 
$E_l = E_l(E)$ is the muon energy at the length $l$ with $E_0(E) = E$,
$E_\phi$ is the energy of the produced $\phi$, $\theta_\phi$ is the angle of the produced $\phi$,
and $\theta_\mathrm{det}(l)$ is the angle between the interaction point and the detector
which we specify below.
We estimate the differential cross section $d^2\sigma/dE_\phi d\cos\theta_\phi$
based on the equivalent photon approximation (EPA) following~\cite{Bjorken:2009mm} 
(see App.~\ref{app:EPA} for more details).
By changing the integration variable from $l$ to $E_l$, we can rewrite the above equation as
\begin{align}
	\Phi_\phi(E_\phi) &= \int dE\,\Phi_\mu (E) \int_{E_\mathrm{min}}^{E} dE_l\,\frac{n_A}{-dE/dl} 
	\int_{0}^{\theta_\mathrm{det}(l(E, E_l))}d\theta_\phi \sin\theta_\phi 
	\frac{d^2\sigma}{dE_\phi d\cos\theta_\phi} (E_l, E_\phi),
	\label{eq:phi_flux_muon_dump}
\end{align}
where
\begin{align}
	l(E, E_l) = \int_{E_l}^{E}\frac{dE_l}{-dE/dl}.
\end{align}
If a muon stops inside the shield $E_\mathrm{min} = 0$, 
otherwise, $E_\mathrm{min}$ is the energy of the muon at the end of the shield.
We read off the stopping power of muon, $-dE/dl$, from~\cite{ParticleDataGroup:2020ssz}.
The angular acceptance is essential for the NuTeV experiment
in which the detector is located far from the muon production (\emph{i.e.} the meson decay) region, and the angular acceptance is of order $10^{-2}\hyphen 10^{-3}$.\footnote{
	This acceptance suppression is due to the fact that the region that dominates the cross section has
	$E_\mu^2 \theta_\phi^2 \lesssim m_\tau^2$.
	We cannot calculate the angular acceptance precisely since 
	we use the EPA that is valid only for $\theta \ll 1$,
	and thus, we put only an order estimation in \autoref{tab:taumuproduction}.
}

NuTeV provides the (anti-)neutrino flux per $10^6$ POT in~\cite{NuTeV:2005wsg},
and they collected $1.13\times 10^{18}$\,POT for positive meson mode
and $1.41\times 10^{18}$\,POT for negative meson mode.
Recognizing that the high energy region of this flux dominantly comes from Kaon decays 
where muon mass is less important, we may estimate the muon flux simply as $\Phi_{\mu^+} = \Phi_\nu$
and $\Phi_{\mu^-} = \Phi_{\bar{\nu}}$.
The shield length of NuTeV is $l_\mathrm{shield} = 850\,\mathrm{m}$.
We take the mass density of the shield as $\rho_A = 2.7\,\mathrm{g}/\mathrm{cm}^3$ 
which is the standard crust density.
We assume for simplicity that the shield is solely composed of Si 
(which is the dominant component of the Earth),
from which we obtain in $n_A \simeq 5.8\times 10^{22}\,\mathrm{cm}^{-3}$.
In this case, we can simply set $E_\mathrm{min} = 0$ as even the highest energy muon 
at NuTeV is well stopped by the shield, as it should be.
The fiducial volume of the decay channel region is $2.54\,\mathrm{m}\times 2.54\,\mathrm{m}\times 34\,\mathrm{m}$~\cite{NuTeV:2000ehk}.
Since the detector is right behind the shield, we have
\begin{align}
	\theta_\mathrm{det}(l) = \frac{1.27\,\mathrm{m}}{850\,\mathrm{m} - l}.
\end{align}
Note that the (anti-)neutrino flux in~\cite{NuTeV:2005wsg}
is normalized by $3\,\mathrm{m}\times 3\,\mathrm{m}$
which is the area of the Lab~E detector,\footnote{
Although they did not explicitly mention, we believe that the flux is normalized by $3\,\mathrm{m}\times 3\,\mathrm{m}$ because 
they are interested in the (anti-)neutrino flux that goes through the Lab~E detector whose area is $9\,\mathrm{m}^2$.
} and thus, we normalize our estimation of the muon flux by
$2.54^2/3^2$ to be conservative.\footnote{
	As we argued above, the angular spread of the produced $\phi$ is relevant in our case.
	This means that the muon flux outside the area of the decay channel region can also contribute
	to the events. 
	Also in this sense our treatment is conservative as we ignore muons outside 
	the detector acceptance cone producing a $\phi$ that is accepted by the detector.
} With this information and Eq.~\eqref{eq:phi_flux_muon_dump} 
we can estimate the flux of $\phi$ at NuTeV.

SHiP provides their expected muon flux in~\cite{SHiP:2020hyy}, and we use it for our study,
with the total number of $2\times 10^{20}$ POT assumed.
The SHiP configuration is described in~\cite{SHiP:2015vad}.
After the target, there is 5\,m of ironic hadron absorber.
Then there is 48\,m of active muon shield by magnetic fields,
followed by 10\,m of $\nu_\tau$ detector.
After the $\nu_\tau$ detector there exists $l_\mathrm{det} = 64\,\mathrm{m}$ 
of decay volume followed by detectors.
In our case, the muon flux produced at the target goes through the hadronic absorber and produces $\phi$
before shielded by the active magnetic field.
Therefore we take $l_\mathrm{shield} = 5\,\mathrm{m}$. 
The produced $\phi$ travels towards the detector and decays
in the decay volume, leaving signals in the detector.
We take the iron mass density $\rho_A = 7.8\,\mathrm{g}/\mathrm{cm}^3$
which results in $n_A = 8.4\times 10^{22}\,\mathrm{cm}^{-3}$.
In this case, the muon energy loss inside the hadron absorber is $\sim 10\,\mathrm{GeV}$
which is small compared to the initial muon energy.
Therefore we may simply ignore the muon energy loss inside the hadron absorber.
The formula is then simplified as
\begin{align}
	\Phi_\phi(E_\phi) &= l_\mathrm{shield} n_A \int dE\,\Phi_\mu (E)
	\int_{0}^{\theta_\mathrm{det}}d\theta_\phi \sin\theta_\phi 
	\frac{d^2\sigma}{dE_\phi d\cos\theta_\phi} (E, E_\phi),
	\label{eq:phi_flux_muon_dump_SHiP}
\end{align}
where we take $\theta_\mathrm{det} = (10+5)/4/(48 + 10 + 64) \simeq 0.031$.
We have checked that, as long as $\theta_\mathrm{det} \gtrsim 0.02$, the angular acceptance
does not affect the flux of $\phi$.

In principle, CHARM can also produce $\phi$ from the $\mu$OT.
However, we could not find an appropriate reference to
estimate the flux of $\mu$ that passes through 
the dirt region between the production point and the detector.
Moreover, the $\mu$OT constraint of CHARM is expected to be at most comparable to NuTeV 
and is well above the couplings relevant for the muon $g-2$ anomaly.
Therefore we do not consider the $\mu$OT production of CHARM in this study.

\subsection{Constraints and future prospects}

\begin{figure}[t]
\centering
	\includegraphics[width=0.5\linewidth]{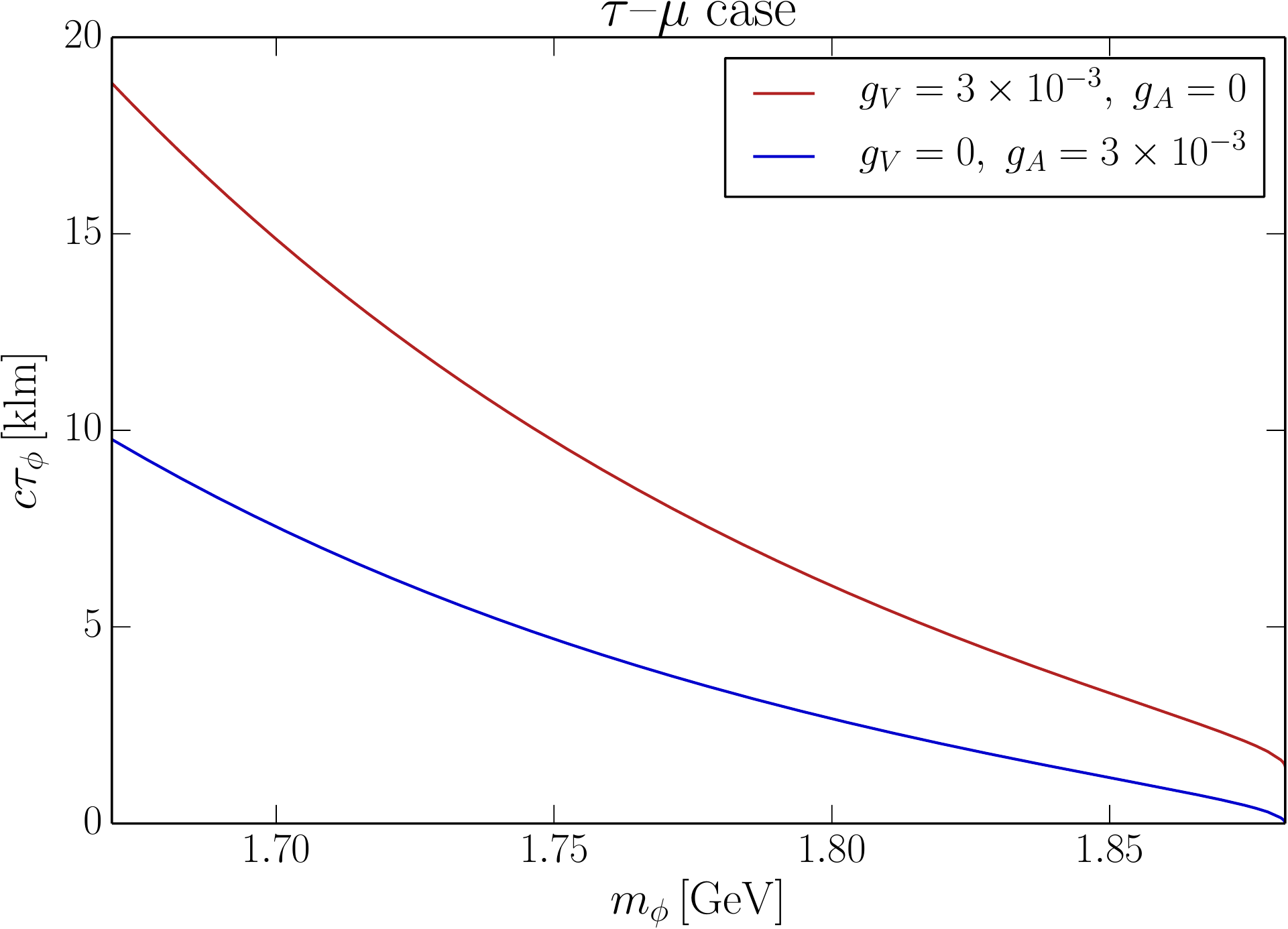}
	\caption{The decay length of $\phi$ in the $\tau-\mu$ case.
	The red line corresponds to $g_V = 3\times 10^{-3}$ and $g_A = 0$,
	while the blue line does to $g_V = 0$ and $g_A = 3\times 10^{-3}$.}
\label{fig:tau_mu_phi_decay}
\end{figure}

 We now derive the constraints on the couplings from the CHARM and NuTeV experiments,
 and the future sensitivity of the SHiP experiment.
 As discussed above, in these experiments, $\phi$ is generated 
 by the direct electroweak production, the $D_s$ decay, and the $\mu$OT processes.
 After production, a fraction of them decay inside the detector and can be identified as signals.
 The event number is estimated as
\begin{align}
	N_{\mu X} &= \int dE_\phi\, \Phi_\phi (E_\phi) 
	\times \frac{l_\mathrm{det}}{\gamma \beta c \tau_\phi},
	\quad
	\frac{1}{\gamma \beta} = \left(\frac{E_\phi^2}{m_\phi^2}-1\right)^{-1/2},
\end{align}
where we assume that the decay length of $\phi$ satisfies $\gamma \beta c\tau_\phi \gg l_\mathrm{det}$.\footnote{
This assumption can be violated for $g \sim 0.1$.
However, this parameter range is out of our interest since it gives a too large contribution to the muon $g-2$,
and thus we ignore this subtlety.
}
In the $\tau\hyphen \mu$ case the computation of $\phi$'s decay rate
is more involved than the $\mu\hyphen e$ case. 
In SM, the branching ratio of $\tau^- \to \mu^- \bar{\nu}_\mu \nu_\tau$ is around 17\,\%.
Therefore, we estimate the decay rate of $\phi$ by assuming that it shares the same branching ratio 17\,\%
for the channel $\phi \rightarrow \mu^- \mu^+ \nu_\mu \bar{\nu}_\tau$,
whose decay width is given by replacing $\mu \to \tau$ and $e \to \mu$ in 
Eq.~\eqref{eq:Gamma_phi_mue}.\footnote{
    Here we focus on the decay with an off-shell $\tau$.
    For $m_\tau + m_e < m_\phi < m_\tau + m_\mu$, $\phi$ can decay as $\phi \to \tau e \nu_e \nu_\mu$
    with an off-shell $\mu$, but this process is well suppressed by the phase space.
    Therefore we ignore this process in the following.
} The intermediate $\tau$ is indeed close to on-shell as the mass difference between $\tau$ and $\phi$
is small, which makes our estimation reasonable.
Note that the decay product of $\phi$ always contains at least one muon, irrespective of the decay
mode of the off-shell $\tau$. For this reason, we denote the event number as $N_{\mu X}$
with $X$ representing the decay product of the off-shell $\tau$.
In \autoref{fig:tau_mu_phi_decay}, we plot the decay length of $\phi$ as a function of $m_\phi$
estimated in this way.
The detector configurations are summarized in \autoref{tab:taumuproduction}.

In order to derive the constraint from NuTeV, we use the analysis in~\cite{NuTeV:2001ndo}.
The NuTeV collaboration detected three $\mu\mu$ events in the decay channel in this analysis.
This is significantly more than the expected background, and the origin of these events is still not clear.
In the current paper, 
we use the CLs method to derive the 95\% C.L. exclusion on the number of signal events in the $\mu\mu$ channel (with the $\mu\mu\nu\bar \nu$ branching fraction of $17\%$) \emph{i.e.},
\begin{align}
	N_{\mu X}(\mathrm{NuTeV}) < 46, 
\end{align}
to estimate the constraint.
Similarly, for the CHARM experiment in the dilepton plus missing energy channel~\cite{CHARM:1985nku}, 2 events were observed with $1.4\pm0.4$ expected background. Hence we have\footnote{The dilepton plus missing energy branching fraction for $\phi$ is similar to $\tau$ leptonic decay branching fraction, which is $35\%$.}
\begin{align}
	N_{\mu X}(\mathrm{CHARM}) < 14.
\end{align}

For the SHiP experiment, as an estimation of its sensitivity, we simply assume that 
our event is background-free.
We then draw the sensitivity line of the SHiP experiment by requiring that
\begin{align}
	N_{\mu X}(\mathrm{SHiP}) < 3.0.
\end{align}
Here we assume all decay channels of $\phi$ can be detected at SHiP experiment with negligible background.

\begin{figure}[t]
    \centering
    \includegraphics[width=0.6\linewidth]{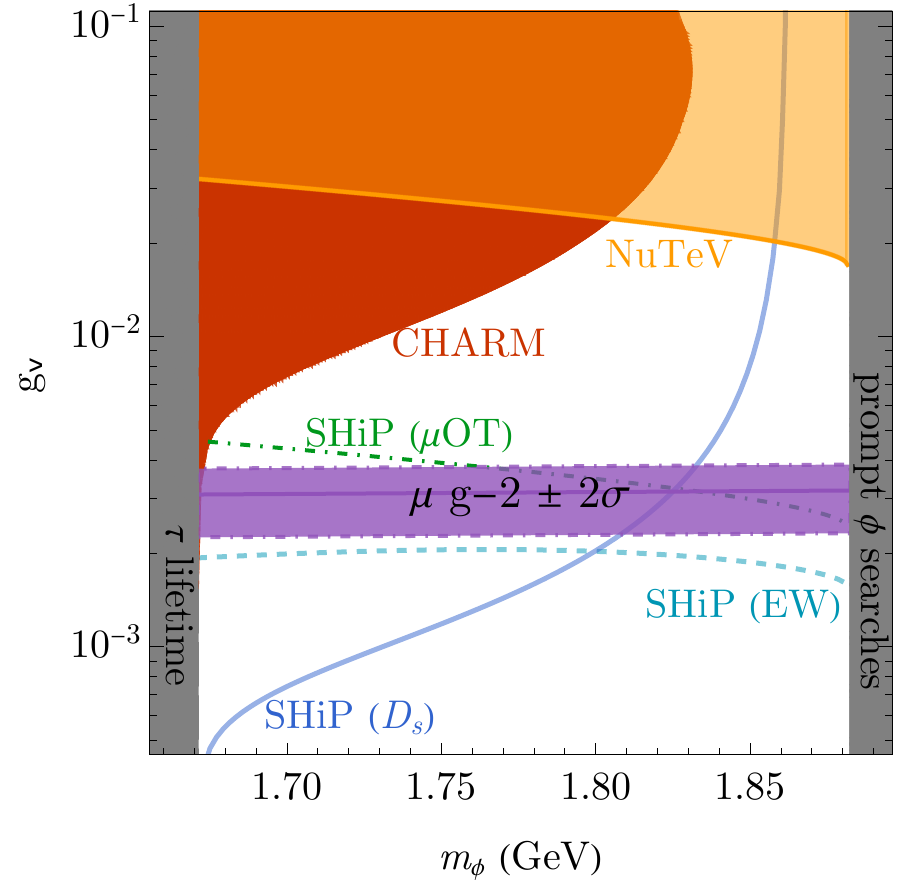}
    \caption{The existing constraint and projection sensitives in the $g_V - m_\phi$ plane for different experiments.
    }
    \label{fig:mu-tau-final}
\end{figure}

We summarize the constraints on $g_V$ as functions of $m_\phi$ in \autoref{fig:mu-tau-final}. 
The purple region is the $2\sigma$ parameter space for the muon $g-2$ anomaly. 
The red and blue regions are covered by the $D_s^\pm$ decay at the CHARM and SHiP, respectively. The green region can be probed from the SHiP electroweak production channel. The orange and blue lines are from the $\mu$OT at NuTeV and SHiP. 
For the three production mechanisms at SHiP, the electroweak production dominates for larger $m_\phi$ while $D_s^\pm$ decay wins for smaller $m_\phi$. 
A large $m_\phi$ suppresses the phase space of the $D_s$ decay as seen in \autoref{fig:Ds_decay_br}. 
On the other hand, the electroweak production is insensitive to the change of $m_\phi$ within the range of our interest. The event number increases as $m_\phi$ increases since a larger $m_\phi$ leads to a smaller decay length that enhances the decay probability of $\phi$ inside the detector.
Here we show the projected sensitivities from the three production mechanisms separately, but of course, we observe
the sum of them in reality, which strengthens our projection.

We can see that CHARM and NuTeV cannot exclude the parameter region that explains the muon $g-2$ anomaly, while the future SHiP experiment covers the whole parameter space. In other words, muon $g-2$ provides the most superior probe of this model, compared to the sensitivity of all the existing beam dump experiments. 
Aside from the big increase in intensity, an essential advantage of the SHiP is that the detector will be placed much closer to the beam dump target than CHARM and NuTeV, thus avoiding suppression from angular acceptance.

Before closing this section, we may comment on the $\tau$ lepton flavor violation (LFV) search at Belle~II.
In~\cite{Tenchini:2020njf}, Belle~II looks for an LFV decay $\tau \to e \alpha$ with $\alpha$ a new particle by 
studying the spectrum of the electron in the $\tau$ pseudo-rest frame constructed from
the decay products of the other $\tau$, following the strategy of ARGUS experiment~\cite{ARGUS:1995bjh}.
A similar technique is applicable to our $\tau\hyphen\mu$ case as the new $\tau$ decay mode
prefers $\phi$ to be close to on-shell, which results in a peak-like structure of the muon spectrum
in the $\tau$ (pseudo-)rest frame. Our estimation indicates that the current limit is not strong
enough to probe the parameter region relevant for the muon $g-2$, mainly because the inferred signal acceptance rate is low, but it would be interesting to
study future Belle~II sensitivity to this channel in more detail.

\section{Summary}
\label{sec:summary}

In this paper, we studied the phenomenology of a complex scalar field $\phi$ that couples dominantly
to $\mu$ and either $e$ or $\tau$, motivated by the muon $g-2$ anomaly.
We focused on the mass range $\abs{m_l - m_\mu} < m_\phi < m_l + m_\mu$ where $l = e$ or $\tau$.
Despite the limited parameter region, this mass range provides us 
with interesting phenomenological signatures. 
The upper bound prohibits the two body-decay $\phi \to l + \mu$, making $\phi$ accidentally long-lived,
while the lower bound prohibits $\tau \to \phi + \mu$ and $\mu \to \phi + e$, avoiding stringent constraints
on the couplings from the lepton flavor violation experiments.

In the $\mu\hyphen e$ case, we found that the whole parameter region that explains the muon $g-2$
anomaly is excluded by the MuLan and LSND experiments.
Our key observation is that, although the mass range $m_\mu - m_e < m_\phi < m_\mu + m_e$
prohibits the decay of free muon $\mu \to \phi + e$, it still allows the decay of muonium
(the bound state of $\mu^+$ and $e^-$) into $\phi$ and photon.
The MuLan experiment performed a dedicated measurement of the muon lifetime with two different materials,
one with the muonium formation and the other without,
from which one can derive a constraint on the muonium invisible decay rate.
More interestingly, $\phi$ can be produced from the muonium decay in neutrino experiments that use
stopped $\mu^+$, such as the LSND experiment.
A fraction of the produced $\phi$ decays inside the LSND detector, from which we constrained
the coupling between $\phi$ and the leptons.

In the $\mu\hyphen \tau$ case, we studied both the past and future high-energy beam dump experiments,
CHARM, NuTeV, and SHiP.
We found that, although CHARM and NuTeV do not exclude the parameter region relevant for the muon $g-2$ anomaly, the future SHiP experiment can cover the entire region.
In the $\mu \hyphen \tau$ case, $\phi$ can be produced from three distinct processes, 
the decay of $D_s$ meson, the electroweak interaction 
and the high-energy muons passing through dense matter.
In particular, we pointed out that the NuTeV experiment can be interpreted as 
a ``muon on target" experiment as the muons produce $\phi$ 
while passing through 850\,m of dirt between the meson decay region and the detector.
This channel covers the large mass end of $m_\phi$ that is not covered by
the $D_s$ decay due to the phase space suppression.
Although SHiP can also be interpreted as the muon beam dump experiment,
the $D_s$ decay and the electroweak production are more efficient as the muon on target is limited 
by the length of the absorber.

Throughout this paper, we focused on the flavor off-diagonal couplings of $\phi$ 
and ignored the flavor diagonal couplings by assigning the global lepton U(1) charge to $\phi$.
This global symmetry is broken due to the neutrino oscillation, but the effect is suppressed 
by the small neutrino mass (difference) and hence is negligible.
Thus, as long as a UV model does not introduce any additional breaking of this (approximate) symmetry,
our assumption that $\phi$ has only the flavor off-diagonal couplings is justified.

Beyond detecting the decays of $\phi$, we can also consider the scattering of $\phi$ with detector materials. The signal and background considerations will require more detailed simulations of the detector performance, which we save for future work. To close up, we would like to point out that the longevity mass windows, identified in this work, can also exist for the coupling with quarks. 
For example, a scalar $\phi$ coupled to $s-d$ current in the $m_{K}-m_\pi < m_\phi < m_{K}+m_\pi $ mass window will lead to  long lifetimes even for a sizable coupling, as the two-body decays to mesons will be forbidden. Such models will lead to intricate phenomenology of $\phi$ production and decay, and for certain mass ranges may add additional rare kaon decay modes, such as $K_L\to \phi \gamma \gamma$ that may compete with the SM $K_L \to \nu\bar \nu 
\pi^0 \to \nu\bar\nu \gamma \gamma $ decay (see {\em e.g.} \cite{Hostert:2020gou}). An in-depth analysis of the long-lived scalars coupled with quarks may deserve further studies. 

\section*{Acknowledgments}
We would like to thank Dr. O Ruchayskiy for a helpful discussion. 
Y.E. and M.P. were supported in part by the DOE grant DE-SC0011842.
Z.L. and K.F.L. were supported in part by the DOE grant DE-SC0022345. Z.L. would like to acknowledge Aspen Center for Physics for hospitality, which is supported by National Science Foundation grant PHY-1607611.
The Feynman diagrams in this paper are generated by \texttt{TikZ-Feynman}~\cite{Ellis:2016jkw}. Z.L. acknowledges the Minnesota Supercomputing Institute (MSI) at the University of Minnesota for providing resources that contributed to the research results reported within this paper \href{http://www.msi.umn.edu}{http://www.msi.umn.edu}.

\appendix

\section{Computational details}

\subsection{$\phi$ decay rate}
\label{app:phi_decay}

In this appendix we explain the computation of $\phi$ decay rate
that is crucial for our study in the main text.
We focus on the $l=e$ case in the following.
As we explained in the main text, the decay rate in the $l = \tau$ case can be reasonably estimated
from the $l = e$ case by replacing $\mu \to \tau$ and $e \to \mu$ in the final result 
and rescaling by the branching ratio of $\tau$.

Since $\mu$ almost always decays into $e \nu_\mu \nu_e$, 
we focus on the decay $\phi \to e^+ e^- \nu_\mu \bar{\nu}_e$ mediated by muon. 
The amplitude is diagrammatically given by
\begin{align}
	i\mathcal{M} = \begin{tikzpicture}[baseline=(v1)]
	\begin{feynman}[inline = (base.v1)]
		\vertex [label=\({\scriptstyle \phi,\,P}\)] (phi);
		\vertex [right = of phi] (v1);
		\vertex [above right = of v1, label=360:\({\scriptstyle e^+,\,p_1}\)] (ep);
		\vertex [below right = of v1] (v2);
		\vertex [above right = of v2, label=360:\({\scriptstyle e^-,\,p_2}\)] (em);
		\vertex [right = of v2, label=360:\({\scriptstyle \nu_\mu,\,p_3}\)] (numu);
		\vertex [below right = of v2, label=360:\({\scriptstyle \bar{\nu}_e,\,p_4}\)] (nue);
		\node [left = -0.05cm of v2, dot,] (fermi);
		\begin{pgfonlayer}{bg}
		\diagram*{
		(phi) -- [charged scalar] (v1) -- (ep),
		(v1) -- [fermion] (v2) -- [fermion] (numu),
		(v2) --(em), (v2) -- (nue),
		};
		\end{pgfonlayer}
	\end{feynman}
	\end{tikzpicture},
\end{align}
where the black dot indicates the 4-fermi interaction
\begin{align}
	\mathcal{L} = -\frac{G_F}{\sqrt{2}}\left[\bar{\nu}_\mu \gamma_\alpha (1-\gamma_5)\mu\right]
	\left[\bar{e}\gamma^\alpha (1-\gamma_5) \nu_e\right],
\end{align}
and the arrow indicates the flow of the muon number.
The amplitude squared is given by
\begin{align}
	\abs{\mathcal{M}}^2 
	&= \frac{128G_F^2}{(p_{234}^2-m_\mu^2)^2} (p_2 \cdot p_3) p_4^\alpha
	\nonumber \\
	&\times \left[
	\abs{g_V +g_A}^2 \left(2(p_1\cdot p_{234}) p_{234 \alpha} - p_{234}^2p_{1\alpha}\right)
	+ \abs{g_V - g_A}^2 m_\mu^2 p_{1\alpha}
	- 2 (\abs{g_V}^2 - \abs{g_A}^2) m_e m_\mu p_{234 \alpha}
	\right],
\end{align}
where we use the notation $p_{ij} = p_i + p_j$ and $p_{ijk} = p_i + p_j + p_k$,
and we take the spin sum of the final particles. The decay rate is given by
\begin{align}
	\Gamma_\phi = \frac{1}{2m_\phi}\int d\Phi_4 \abs{M}^2.
	\label{eq:Gamma_4body}
\end{align}
We may decompose the phase space integral as
\begin{align}
	\int d\Phi_4
	&= \int_{m_e^2}^{(m_\phi-m_e)^2}\frac{d\bar{m}_\mu^2}{2\pi}
	\int d\Phi_{1;234}
	\int_{0}^{(\bar{m}_\mu-m_e)^2}\frac{d\bar{m}_{\nu\nu}^2}{2\pi}
	\int d\Phi_{2;34}\int d\Phi_{34},
\end{align}
where 
\begin{align}
	\int d\Phi_{1;234}
	&= \int \frac{d^3p_1}{(2\pi)^3 2E_1} \int \frac{d^3 p_{234}}{(2\pi)^3 2\bar{E}_{234}}
	(2\pi)^4 \delta^{(4)}(P-p_1 - p_{234}), \\
	\int d\Phi_{2;34}
	&= \int\frac{d^3 p_2}{(2\pi)^32E_2}
	\int \frac{d^3 p_{34}}{(2\pi)^3 2 \bar{E}_{34}}(2\pi)^4 \delta^{(4)}(p_{234}-p_2 - p_{34}),
	\\
	\int d\Phi_{34}
	&= \int\frac{d^3 p_3}{(2\pi)^3 2 p_3}\int \frac{d^3 p_4}{(2\pi)^3 2p_4}
	(2\pi)^4 \delta^{(4)}(p_{34}-p_3 - p_4),
\end{align}
and we denote $p_{234}^2 = \bar{m}_\mu^2$, $p_{34}^2 = \bar{m}_{\nu\nu}^2$,
$\bar{E}_{234}^2 = {\abs{\vec{p}_{234}}^2 + \bar{m}_\mu^2}$, and
$\bar{E}_{34}^2 = {\abs{\vec{p}_{34}}^2 + \bar{m}_{\nu\nu}^2}$.
We first note that the matrix element squared depends on $p_3$ and $p_4$ only
through the combination $p_3^\alpha p_4^\beta$.
After the phase space integral we find that
\begin{align}
	\int d\Phi_{34}\,p_3^\alpha p_4^\beta
	= \frac{1}{96\pi}\left(\eta^{\alpha\beta}\bar{m}_{\nu\nu}^2 + 2p_{34}^\alpha p_{34}^\beta\right),
\end{align}
by evaluating the integral in the $p_{34}$-rest frame and exploiting the Lorentz invariance.
Therefore we obtain
\begin{align}
	\int d\Phi_{2;34} \int d\Phi_{3;4}
	(p_2 \cdot p_3)p_4^\alpha
	&= \frac{1}{96\pi}\int d\Phi_{2;34} \left[p_2^\alpha \bar{m}_{\nu\nu}^2 + 2(p_2 \cdot p_{34}) p_{34}^\alpha\right]
	= \frac{I_{1}(\bar{m}_\mu, \bar{m}_{\nu\nu})}{1536\pi^2}p_{234}^\alpha,
\end{align}
where we use the Lorentz invariance in the last line. 
By working in the $p_{234}$-rest frame, we find that
\begin{align}
	I_{1} &= \left[1+\frac{\bar{m}_{\nu\nu}^2-2m_e^2}{\bar{m}_\mu^2}
	+ \frac{m_e^4 + m_e^2 \bar{m}_{\nu\nu}^2 - 2\bar{m}_{\nu\nu}^4}{\bar{m}_\mu^4}
	\right]
	\sqrt{\left(\bar{m}_\mu^2 - \bar{m}_{\nu\nu}^2\right)^2
	- 2m_e^2 (\bar{m}_\mu^2 + \bar{m}_{\nu\nu}^2) + m_e^4}.
\end{align}
By substituting this to Eq.~\eqref{eq:Gamma_4body} and performing the angular integral,
we obtain Eq.~\eqref{eq:Gamma_phi_mue}.

\subsection{Muonium decay rate}
\label{app:muonium_decay}

Since we take $m_\mu - m_e < m_\phi < m_\mu + m_e$,
muonium decay to $\phi$ and $\gamma$ is kinematically allowed.
We compute the branching ratio of this process here.
The muonium state is written as (see e.g.~\cite{Peskin:1995ev})
\begin{align}
	\left\vert \mathrm{Mu}\right\rangle = \sqrt{2E_{\mathrm{Mu}}}
	\int\frac{d^3 k}{(2\pi)^3}\tilde{\psi}(\vec{k}) \frac{1}{\sqrt{2m_\mu}}\frac{1}{\sqrt{2m_e}}
	\lvert \mu; \vec{k}\rangle \otimes \lvert e; -\vec{k}\rangle,
\end{align}
where $E_\mathrm{Mu}$ is the energy of the muonium 
and $\tilde{\psi}$ is the (momentum-space) muonium wavefunction.
The amplitude is then given by
\begin{align}
	\mathcal{M}(\mathrm{Mu}\to \phi \gamma)
	= \sqrt{2E_{\mathrm{Mu}}}
	\int\frac{d^3 k}{(2\pi)^3}\tilde{\psi}(\vec{k}) \frac{1}{\sqrt{2m_\mu}}\frac{1}{\sqrt{2m_e}}
	\mathcal{M}_\mathrm{free}(\mu e \to \phi \gamma),
\end{align}
where $\mathcal{M}_\mathrm{free}$ is the matrix element between free particles.
We work in the muonium rest frame.
Inside muonium both muon and electron are non-relativistic ($v \sim \alpha \ll 1$),
and thus we work only on the leading order in the non-relativistic limit,
dropping $\vec{k}$ dependence of $\mathcal{M}_\mathrm{free}$.
We also work in the limit $m_e/m_\mu \ll 1$ and keep only the leading order terms.
We then obtain
\begin{align}
	\Gamma(\mathrm{Mu}\to \phi\gamma) &= \frac{\vert \vec{p}\vert \abs{\psi(0)}^2}{64\pi^2 m_\mu^2m_e}
	\int d\Omega_2 \abs{\mathcal{M}_\mathrm{free}(\mu e \to \phi \gamma)}^2,
\end{align}
where $\vec{p}$ is the momentum of the final state photon,
$\psi(0)$ is the coordinate-space muonium wavefunction at the origin,
and we ignore the binding energy which is of order $\alpha^2 m_e$.
Note that this can be written as
\begin{align}
	\Gamma(\mathrm{Mu}\to \phi\gamma) &= \abs{\psi(0)}^2 \times \sigma_\mathrm{free} v(\mu e \to \phi \gamma).
\end{align}
We may parametrize $m_\phi$ as
\begin{align}
	m_\phi = m_\mu + x m_e,
\end{align}
where $-1 < x < 1$ in the case of our interest. The kinematics then tells us that
\begin{align}
	\abs{\vec{p}} \simeq m_e (1-x),
	\quad
	E_\phi \simeq m_\mu + xm_e.
\end{align}
If we take the ground state muonium, the wavefunction at the origin is given by
\begin{align}
	\abs{\psi(0)}^2 = \frac{\alpha^3 m_e^3}{\pi}.
\end{align}
We thus obtain
\begin{align}
	\Gamma(\mathrm{Mu}\to \phi\gamma) &= \frac{(1-x)\alpha^3}{64\pi^3}\frac{m_e^3}{m_\mu^2}
	\int d\Omega_2 \abs{\mathcal{M}_\mathrm{free}(\mu e \to \phi \gamma)}^2.
\end{align}
Now comes the evaluation of the matrix element squared in the non-relativistic limit.
For this it is convenient to work in the Dirac representation
\begin{align}
	\gamma^0 = \begin{pmatrix} 1 & 0 \\ 0 & -1 \end{pmatrix},
	\quad
	\gamma^i =  \begin{pmatrix} 0 & \sigma^i \\ -\sigma^i & 0 \end{pmatrix},
	\quad
	\gamma_5 =  \begin{pmatrix} 0 & 1 \\ 1 & 0 \end{pmatrix}.
\end{align}
In this representation the spinors in the non-relativistic limit are given by
\begin{align}
	u_e = \sqrt{2m_e}\begin{pmatrix} \chi_{e^-} \\ 0 \end{pmatrix},
	\quad
	v_\mu = \sqrt{2m_\mu}\begin{pmatrix} 0 \\ \chi_{\mu^+} \end{pmatrix}.
\end{align}
We also note that
\begin{align}
	(p_\mu - p_\phi)^2 - m_e^2 \simeq -2(1-x)m_e^2,
	\quad
	(p_\mu - p_\gamma)^2 - m_\mu^2 \simeq -2m_\mu m_e (1-x).
\end{align}
We thus obtain after some computation that
\begin{align}
	\mathcal{M}_\mathrm{free} \simeq
	e \sqrt{\frac{m_\mu}{m_e}}\left[g_Vn^i + ig_A \epsilon^{ijk}\hat{p}^j n^k\right]\left(\epsilon^i_\lambda\right)^*,
	\quad
	\hat{p}^i = \frac{p^i}{\vert\vec{p}\vert},
	\quad
	n^i = \chi_{\mu^+}^\dagger \sigma^i \chi_{e^-},
\end{align}
where $\vec{\epsilon}_\lambda$ is the polarization vector of the photon with its polarization $\lambda$.
We note in passing that only the $t$-channel process gives a finite contribution above.
After taking the polarization sum of the photon, the matrix element squared is given by
\begin{align}
	\sum_\lambda \abs{M_\mathrm{free}}^2
	= e^2 \frac{m_\mu}{m_e}\left(\abs{g_V}^2 + \abs{g_A}^2\right)
	\left(\abs{\vec{n}}^2 - \abs{\hat{p}\cdot \vec{n}}^2\right).
\end{align}
The direction of $\vec{p}$ is unrelated to $\vec{n}$, and thus after the solid angle integral
we obtain
\begin{align}
	\Gamma(\mathrm{Mu}\to \phi^*\gamma) &=
	\frac{1-x}{6\pi}\frac{\alpha^4 m_e^2}{m_\mu}\left(\abs{g_V}^2 + \abs{g_A}^2\right) \abs{\vec{n}}^2.
\end{align}
In order to evaluate $\abs{\vec{n}}^2$, we may note that
\begin{align}
	\chi_{\mu^+} &= \begin{pmatrix} 0 \\ 1 \end{pmatrix}
	~~\mathrm{for}~~ \mu^+: \vert\uparrow \rangle,
	\quad
	\chi_{\mu^+} = \begin{pmatrix} -1 \\ 0 \end{pmatrix}
	~~\mathrm{for}~~ \mu^+: \vert\downarrow \rangle,
\end{align}
since $\mu^+$ is anti-particle.
It then follows that $\abs{\vec{n}}^2 = 2$ for triplet and $\abs{\vec{n}}^2 = 0$ for singlet.
We thus have
\begin{align}
	\Gamma\left(\mathrm{Mu}^{(S=1)}\to \phi^*\gamma\right) &=
	\frac{1-x}{3\pi}\frac{\alpha^4 m_e^2}{m_\mu}\left(\abs{g_V}^2 + \abs{g_A}^2\right),
	\quad
	\Gamma\left(\mathrm{Mu}^{(S=0)}\to \phi^*\gamma\right) = 0,
\end{align}
reproducing Eqs.~\eqref{eq:muonium_decay1} and~\eqref{eq:muonium_decay2} in the main text.

\subsection{Equivalent photon approximation}
\label{app:EPA}
In this subsection we explain the EPA that we used to estimate the flux of $\phi$ in the muon beam dump.
Following~\cite{Bjorken:2009mm}, we approximate
the scattering cross section of $\mu(p) N(P_i) \to \tau(p') \phi(k) N(P_f)$ as
\begin{align}
	\frac{d\sigma(p + P_i \to p' + k + P_f)}{dE_\phi d\cos\theta_\phi}
	= \frac{\alpha \chi}{\pi}\frac{E_\mu x \beta_\phi}{1-x}
	\left.\frac{d\sigma(p+q\to p' + k)}{d(p\cdot k)}\right\vert_{t=t_\mathrm{min}},
	\label{eq:EPAwdiff}
\end{align}
where
\begin{align}
	q = P_i - P_f,
	\quad
	x = E_\phi/E_\mu,
	\quad
	t = -q^2,
	\quad
	\beta_\phi = \sqrt{1-\frac{m_\phi^2}{E_\mu^2}},
\end{align}
and $\theta_\phi$ is the scattering angle between the initial muon and the final $\phi$ in the lab frame.
Here $\sigma(p + q \to p' + k)$ represents the cross section of
$\mu(p)  \gamma(q) \to  \tau(p') \phi(k)$ with $\gamma$ indicating the virtual photon
created by the nucleus.
The EPA assumes that
\begin{align}
	m_\tau, m_\phi, m_\mu, \abs{\vec{q}} \ll E_\mu,
	\quad
	\abs{\vec{q}} \ll m_A,
	\quad
	\theta_\phi \ll 1,
\end{align}
where $m_A$ is the mass of the target nucleus.
The above formula assumes that $\vec{q}$ is collinear with $\vec{k}-\vec{p}$ such that
the photon virtuality $t$ takes its minimum $t_\mathrm{min}$ (note that $m_\tau^2 = (p-k+q)^2$ is fixed).
To the leading order in this kinematics, we obtain in the lab frame (where $P_i = (m_A, 0, 0, 0)$)
\begin{align}
	q^0 &\simeq -\frac{\abs{\vec{q}}^2}{2m_A} \ll \abs{\vec{q}},
	\quad
	\abs{\vec{q}} \simeq -\frac{u_2 - m_\tau^2}{2E_\mu(1-x)},
	\quad
	t_\mathrm{min} \simeq \left(\frac{u_2 - m_\tau^2}{2E_\mu(1-x)}\right)^2, \\
	u_2 &= (p-k)^2 \simeq -E_\mu^2 x \theta_\phi^2 - \frac{1-x}{x}m_\phi^2 + (1-x)m_\mu^2, \\
	s_2 &= (p+q)^2 \simeq m_\mu^2 - \frac{u_2-m_\tau^2}{1-x},
	\quad
	t_2 = m_\phi^2 + \frac{x (u_2-m_\tau^2)}{1-x},
\end{align}
where the quantities with the subscript $2$ indicate the Mandelstam variables of $\mu \gamma \to \tau \phi$.
The form factor is given by
\begin{align}
	\chi = \int_{t_\mathrm{min}}^{t_\mathrm{max}}dt\frac{t-t_\mathrm{min}}{t^2}F^2(t),
	\quad
	F^2(t) = \frac{Z^2}{(1+t/d)^2},
\end{align}
with $d = 0.164\,\mathrm{GeV}^2 A^{-2/3}$ and $t_\mathrm{max} = m_\phi^2$. 
We ignore the atomic form factor since that is irrelevant for our energy scale.
Finally the differential cross section of $\mu \gamma \to \tau \phi$ is given by
\begin{align}
	\frac{d\sigma(p+q\to p' + k)}{d(p\cdot k)} = 2\frac{d\sigma}{dt_2}
	\simeq \frac{\abs{\bar{\mathcal{M}}}^2}{8\pi s_2^2},
\end{align}
where 
\begin{align}
	\abs{\bar{\mathcal{M}}}^2
	= -\frac{e^2}{4}
	\mathrm{Tr}&\left[
	\left(\slashed{p}' + m_\tau\right)
	\left(\frac{(g_V^* - g_A^*\gamma_5)(\slashed{p}+\slashed{q}+m_\mu)}{s_2-m_\mu^2}\gamma^\alpha
	+ \gamma^\alpha\frac{(\slashed{p}-\slashed{k} + m_\tau)(g_V^*-g_A^*\gamma_5)}{u_2-m_\tau^2}
	\right)
	\right. \nonumber \\ &\left.
	\times \left(\slashed{p} + m_\mu\right)
	\left(\gamma_\alpha\frac{(\slashed{p}+\slashed{q}+m_\mu)(g_V + g_A\gamma_5)}{s_2-m_\mu^2}
	+ \frac{(g_V+g_A\gamma_5)(\slashed{p}-\slashed{k} + m_\tau)}{u_2-m_\tau^2}\gamma_\alpha
	\right)
	\right].
\end{align}
We set $t_\mathrm{min} = 0$ in the evaluation of this expression.

\bibliographystyle{JHEP}
\bibliography{ref}

\end{document}